\documentclass[11pt]{article}
\usepackage{amsmath,amsthm,amssymb,authblk}
\usepackage{hyperref}
\usepackage[a4paper,margin=2.5cm]{geometry}
\usepackage{cite}
\usepackage{tikz}
\usetikzlibrary{calc,decorations.pathmorphing,decorations.markings,decorations.pathreplacing,patterns,shapes,arrows}

\tikzset{boper/.style={rectangle,fill,inner sep=2pt,black}}
\tikzset{bblob/.style={circle,fill,inner sep=1.5pt,black}}
\tikzset{blob/.style={circle,draw,fill=white,outer sep=0mm,inner sep=0.3mm}}
\tikzstyle directed=[postaction={decorate,decoration={markings,
    mark=at position .65 with {\arrow[arrowstyle]{stealth}}}}]

\pgfdeclaremetadecoration{prewavy}{zigzag}{
  \state{zigzag}[width=\pgfmetadecorationsegmentamplitude*\pgfmetadecoratedpathlength - 0.5*\pgfmetadecorationsegmentlength ,
                  next state=straight] {
        \decoration{zigzag}
      }
    \state{straight}[width=\pgfmetadecorationsegmentlength,
                   next state=final] {
        \decoration{curveto}
    }
    \state{final}{
        \decoration{zigzag}
        \beforedecoration{\pgfpathmoveto{\pgfpointmetadecoratedpathfirst}}
    }
}

\tikzset{
  wavy/.style={
    decorate,
    decoration={
      prewavy,
      meta-amplitude=#1,
      meta-segment length=0.3cm,
      amplitude=1.5pt, 
      segment length=6pt 
},
    postaction={decorate,ultra thick,decoration={markings,mark = at position #1 with {\arrow{>}}}}        
  },
  wavy/.default=0.5
}

\tikzset{
  aline/.style={
    decorate,
    decoration={
      meta-amplitude=#1,
      meta-segment length=0.3cm,
},
    postaction={decorate,ultra thick,decoration={markings,mark = at position #1 with {\arrow{>}}}}        
  },
  aline/.default=0.9
}

\tikzset{
  ddashed/.style={dashed,
    decorate,
    decoration={
      meta-amplitude=#1,
      meta-segment length=0.3cm,
},
    postaction={decorate,ultra thick,decoration={markings,mark = at position #1 with {\arrow{>}}}}        
  },
  ddashed/.default=0.9
}

\tikzset{
  bline/.style={blue,
    decorate,
    decoration={
      meta-amplitude=#1,
      meta-segment length=0.3cm,
},
    postaction={decorate,ultra thick,decoration={markings,mark = at position #1 with {\arrow{>}}}}        
  },
  bline/.default=0.9
}

\tikzset{
  sdash/.style={
    dashed,
    decorate,
    decoration={
      meta-amplitude=#1,
      meta-segment length=0.3cm,
},
    postaction={decorate,ultra thick,decoration={markings,mark = at position #1 with {\arrow{>}}}}        
  },
  sdash/.default=0.9
}

\graphicspath{{figpdf/}}

\makeatletter

\@addtoreset{equation}{section}
\makeatother

\newcommand{\be}{\begin{eqnarray}}
\newcommand{\ee}{\end{eqnarray}}
\newcommand{\ben}{\begin{eqnarray*}}
\newcommand{\een}{\end{eqnarray*}}
\newcommand{\bec}{\begin{equation}\begin{array}{lll}}
\newcommand{\eec}{\end{array}\end{equation}}

\newcommand{\bW}{\overline{W}}

\newcommand{\ol}{\overline}
\newcommand{\C}{\mathbb{C}}
\newcommand{\Z}{\mathbb{Z}}
\newcommand{\cO}{{\cal O}}
\newcommand{\cC}{{\cal C}}

\newcommand{\w}{\omega}
\newcommand{\ep}{\varepsilon}

\newcommand{\id}{\mathbb{I}}

\newcommand{\ot}{\otimes}
\newcommand{\sli}{\sum\limits}
\newcommand{\squad}{\hspace*{1mm}}
\newcommand{\sn}{\hbox{ sn}}
\newcommand{\cn}{\hbox{ cn}}
\newcommand{\dn}{\hbox{ dn}}
\newcommand{\uq}{U_q(\widehat{\mathfrak{sl}}_2)}
\newcommand{\uqt}{\widetilde{U}_q(\widehat{\mathfrak{sl}}_2)}
\newcommand{\nn}{\nonumber}

\newcommand{\figref}[1]{Figure~\ref{fig:#1}}
\newcommand{\secref}[1]{Section~\ref{sec:#1}}
\newcommand{\aver}[1]{\langle #1 \rangle}

\newcommand{\cA}{{\cal A}}
\newcommand{\ra}{\rightarrow}
\newcommand{\thab}[2]{\Theta_{#1}{}^{#2}}
\newcommand{\thabhat}[2]{\widehat\Theta^{#1}{}_{#2}}
\newcommand{\R}{\mathbb{R}}

\tikzset{oper/.style={rectangle,fill,inner sep=2.5pt}}
\tikzset{arr/.style={postaction={decorate,thick,decoration={markings,mark = at position #1 with {\arrow{>}}}}}}

\title{Discrete Holomorphicity in the Chiral Potts Model}

\author[1,2]{Yacine Ikhlef\thanks{\tt ikhlef@lpthe.jussieu.fr}}
\author[3]{Robert Weston\thanks{\tt R.A.Weston@hw.ac.uk}}
\affil[1]{Sorbonne Universit\'es, UPMC Univ Paris 06, UMR 7589, LPTHE, F-75005, Paris, France \bigskip}
\affil[2]{CNRS, UMR 7589, LPTHE, F-75005, Paris, France \bigskip}
\affil[3]{Department of Mathematics, Heriot-Watt University, Edinburgh EH14 4AS, UK,
  and Maxwell Institute for Mathematical Sciences, Edinburgh, U.K.}

\begin{document}
\bibliographystyle{unsrt}




\maketitle

\begin{abstract}
\noindent 
We construct lattice parafermions for the $Z(N)$ chiral Potts model in terms of quasi-local currents of the underlying quantum group. We show that the conservation of the quantum group currents leads to twisted discrete-holomorphicity (DH) conditions for the parafermions. At the critical Fateev-Zamolodchikov point the parafermions are the usual ones, and the DH conditions coincide with those found previously by Rajabpour and Cardy. Away from the critical point, we show that our twisted DH conditions can be understood as deformed lattice current conservation conditions for an underlying perturbed conformal field theory in both the general $N\geq 3$ and $N=2$ Ising cases.

\end{abstract}

\nopagebreak

\section{Introduction}\label{sec:intro}

The chiral Potts model is a two-dimensional statistical model defined by spin variables subject to a $\Z_N$-symmetric, local interaction, and was introduced in the 1980s as a lattice model for commensurate-incommensurate phase transitions~\cite{HKdN83}. A few years later, it became of great interest for mathematical physics as a solution of the star-triangle equations~\cite{star-triangle88} in which the Boltzmann weights do not satisfy the difference property, and also as a superintegrable~\cite{Bax88b} generalisation of the Ising model. An important step in the understanding of the model was its identification~\cite{BazhStrog90} as a {descendent} of the well-studied six-vertex model -- more precisely, the integrable chiral Potts model is based on $\Z_N$ cyclic representations~\cite{Babelon84,RocheArn89} of the affine quantum algebra $\uq$ (see also~\cite{BP90}). This class of representations was then studied in detail and generalised to other quantum algebras in~\cite{DJMM90sl3,date1990new,DJMM91gln}.

The chiral Potts model displays very peculiar physical features. In the superintegrable regime, it describes a commensurate-incommensurate phase transition in an intrinsically aniso\-tropic lattice model~\cite{McCoyRoan90,Alb89a,Alb89b,Alb91}. This behaviour is believed to appear also in the ordinary integrable regime, where the chiral Potts model provides an integrable chiral deformation~\cite{Cardy93} of the Fateev-Zamolodchikov (FZ) clock model~\cite{FZclock}. The latter is an integrable $\Z_N$-symmetric spin model whose scaling limit is described by the $\Z_N$-parafermionic current algebra~\cite{FZcft}, an extension of the Virasoro algebra generating the spectrum of a Conformal Field Theory (CFT). By a simple inspection of the Boltzmann weights, Rajabpour and Cardy have observed~\cite{RajCardy} that one of the parafermionic currents $\{\psi_k\}$ of the $\Z_N$-parafermionic CFT has a lattice analog in the FZ clock model, which is a {\it discretely holomorphic} operator, i.e., it satisfies a discrete version of the Cauchy-Riemann equations. Discretely holomorphic parafermions have been found empirically in a number of critical lattice models~\cite{RivaCardy,RajCardy,IkhCardy,deGierLR13}, and are a crucial ingredient to the rigorous study of bulk~\cite{Smi01,Smi07,ChelkakS09,DuminilS12,ChelkakHI12} and boundary~\cite{BeatondGG11} critical properties. In a recent paper~\cite{IWWZ}, we have shown that the origin of discretely holomorphic parafermions for loop models can be traced to the underlying quantum algebraic structure, following the construction of quasi-local conserved currents by Bernard and Felder~\cite{BF91}.

The object of the present paper is to use the integrability and $\uq$ symmetry of the chiral Potts model to (i) explain the algebraic origin of the discretely holomorphic parafermions observed~\cite{RajCardy} for the FZ clock model, and (ii) extend the discrete Cauchy-Riemann equations to the chiral regime around the FZ point, and analyse their physical meaning in the scaling limit.

The paper is organised as follows. Sections \ref{sec:CP}, \ref{sec:NLQGC} and \ref{sec:CPRT} are devoted to reviewing some background material, respectively on the basics of the chiral Potts model, the Bernard-Felder construction, and the $\uq$ symmetry of the chiral Potts model. In~\secref{NLCCPM} we construct the quasi-local operators associated to generators of $\uq$ (which reduce to the lattice parafermions of~\cite{RajCardy} at the FZ point), and we give an explicit form of the linear relations generalising the discrete Cauchy-Riemann equations for these operators. In \secref{phys} we interpret these linear relations in terms of perturbed CFT~\cite{Zam89} and discuss both the general $N\geq 3$ model
and $N=2$ Ising case.  Finally, we summarise our findings in \secref{conclu}.

\section{The Chiral Potts Model}
\label{sec:CP}


\subsection{Definitions}
The chiral Potts (CP) model \cite{HKdN83,star-triangle88,Bax88b} is a statistical model where the variables are spins $a_j \in \Z_N$ living on the sites of a square lattice $\mathcal{L}$.
The Boltzmann weight of a spin configuration $\{a_j\}$ is invariant under a rotation of all spins $(a_j \to a_j+1 \mod N)$, and is specified by the discrete functions $W_{\aver{ij}}$ associated to the edges of $\mathcal{L}$:
\begin{equation}
  \mathcal{W}[\{a_j\}] = \prod_{\aver{ij}} W_{\aver{ij}}(a_i-a_j) \,,
\end{equation}
where $\aver{ij}$ denotes a pair of neighbouring sites, connected by an oriented edge $i \to j$. Let us describe the specific choice of weight functions $W_{\aver{ij}}$ which renders the model integrable. In addition to the number of spin states $N$, we fix an external real parameter $k' \geq 0$. Each rapidity line carries a spectral parameter $\xi$ given as a triplet of complex numbers $\xi = (x,y,\mu)$ obeying the algebraic equations
\begin{equation} \label{eq:Ck}
  x^N+y^N = k(1+x^Ny^N) \qquad \text{and} \qquad \mu^N(1-kx^N)=k' \,,
\end{equation}
where we have set $k=\sqrt{1-k'^2}$. A SW$\to$NE (resp. NW$\to$SE) edge crossed by rapidity lines $(r,s)$ is assigned the weight function $W_{rs}$ (resp. $\ol W_{rs}$), defined\footnote{
The conditions~\eqref{eq:Ck} ensure that $W_{rs}$ and $\ol W_{rs}$ are well-defined, i.e., $W_{rs}(a+N)=W_{rs}(a)$ and $\ol W_{rs}(a+N)=\ol W_{rs}(a)$.
} for $a \in \{0, 1, 2, \dots \}$ as
\begin{equation} \label{eq:W}
  W_{rs}(a) = \left( \frac{\mu_r}{\mu_s} \right)^a \times \prod_{\ell=1}^a  \frac{y_s-x_r \w^{\ell}}{y_r-x_s \w^{\ell}} \,,
  \squad
  \bW_{rs}(a) = \left(\mu_r\mu_s \right)^a \times \prod_{\ell=1}^a \frac{x_r \w-x_s \w^{\ell}}{y_s-y_r \w^{\ell}} \,,
\end{equation}
where $\w=\exp(2i\pi/N)$, and we have used  $\xi_r=(x_r,y_r,\mu_r)$ and $\xi_s=(x_s,y_s,\mu_s)$ to denote the spectral parameters attached to the rapidity lines $r$ and $s$, respectively.

The CP weights are represented by
\begin{equation}
  W_{rs}(a-b)=
  \begin{tikzpicture}[baseline=-3pt,scale=0.6]
    \draw[aline] (-2,1) node[left] {$r$} --  (0,-1);
    \draw[aline] (0,1) node[left] {$s$} --  (-2,-1);
    \draw[bline] (-2,0) node[left] {$a$} -- (0,0) node[right] {$b$} ;
    \draw (-2,0) node[bblob] {}; \draw (0,0) node[bblob] {};
  \end{tikzpicture},
  \qquad \bW_{rs}(a-b)=
  \begin{tikzpicture}[baseline=-3pt,scale=0.6]
    \draw[aline] (-2,1) node[left] {$r$} --  (0,-1);
    \draw[aline] (0,1) node[left] {$s$} --  (-2,-1);
    \draw[bline] (-1,1) node[above] {$a$} -- (-1,-1) node[below] {$b$} ;
    \draw (-1,-1) node[bblob] {}; \draw (-1,1) node[bblob] {};
  \end{tikzpicture}\quad.
\end{equation}
We sometimes use an alternative graphical notation for CP weights that emphasises the fact that we
can associate them with rhomboids on the covering lattice (i.e. the union of the CP lattice points,
denoted by $\bullet$, and the dual lattice points denoted by by $\circ$). This notation is
\ben W_{rs}(a-b)=
\begin{tikzpicture}[baseline=-3pt,scale=0.6]
\draw[bline] (-2,0) node[left] {$a$} -- (0,0) node[right] {$b$} ;
\draw (-2,0) node[bblob] {}; \draw (0,0) node[bblob] {};
\draw[dashed] (-1,-1) --  (-2,0) -- (-1,1) -- (0,0) -- cycle;
\draw (-1,-1) circle [radius=0.08];\draw (-1,1) circle [radius=0.08];
\end{tikzpicture},\qquad \bW_{rs}(a-b)=
\begin{tikzpicture}[baseline=-3pt,scale=0.6]
\draw[bline] (-1,1) node[above] {$a$} -- (-1,-1) node[below] {$b$} ;
\draw (-1,-1) node[bblob] {}; \draw (-1,1) node[bblob] {};
\draw[dashed] (-1,-1) --  (-2,0) -- (-1,1) -- (0,0) -- cycle;
\draw (-2,0) circle [radius=0.08];\draw (0,0) circle [radius=0.08];
\end{tikzpicture}\quad.
\een

\begin{figure}[h!]\centering
  \begin{tikzpicture}[baseline=-3pt,scale=0.6]
    \draw (4,1) node[below] {$s$} --  (1,-2);
    \draw (4,3) node[right] {$s$} --  (-1,-2);
    \draw (3,4) node[right] {$s$} --  (-2,-1);
    \draw (1,4) node[right] {$ \;r$} --  (-2,1);
    \draw  (-2,1) node[below] {$r$}--(1,-2);
    \draw  (-2,3) node[left] {$r$}--(3,-2);
    \draw  (-1,4) node[left] {$r$}--(4,-1);
    \draw  (1,4) node[left] {$s\;$}--(4,1);
    \draw[bline] (-1,-1) node[bblob] {}-- (1,-1) node[bblob] {} -- (3,-1) node[bblob] {};
    \draw[bline] (-1,1) node[bblob] {}-- (1,1) node[bblob] {} -- (3,1) node[bblob] {};
    \draw[bline] (-1,3) node[bblob] {}-- (1,3) node[bblob] {} -- (3,3) node[bblob] {};
    \draw[bline] (-1,3) -- (-1,1) --(-1,-1) ;
    \draw[bline] (1,3) -- (1,1) --(1,-1) ;
    \draw[bline] (3,3) -- (3,1) --(3,-1) ;
  \end{tikzpicture}
  \caption{A homogenous Chiral Potts Lattice}\label{fig:CPlattice}
\end{figure}
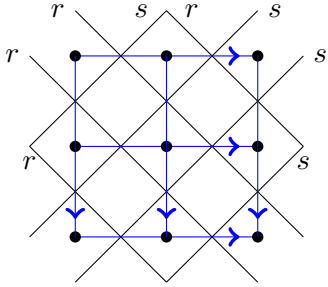

A homogeneous chiral Potts model partition function $Z_{rs}$ is given as the sum over the height variables (i.e., the indices $a\in\{0,1,\cdots,N-1\}$ at the positions marked by $\bullet$) associated with a lattice with spectral parameters  $\xi_r,\xi_s$ distributed as shown in \figref{CPlattice}, in which all diagonal lines have downward arrows, which we have omitted for clarity.

The above weight functions satisfy~\cite{star-triangle88} the star-triangle equations:
\begin{equation} \label{eq:star-triangle}
  \sum_{d=0}^{N-1} \ol W_{rs}(a-d) W_{rt}(d-b) \ol W_{st}(d-c)
  = \rho_{rst} \times W_{rs}(c-b) \ol W_{rt}(a-c) W_{st}(a-b) \,,
\end{equation}
for any fixed spins $(a,b,c)$, and the overall factor $\rho_{rst}$ is a function of $(\xi_r,\xi_s,\xi_t)$ .

\subsection{Crossing symmetry}

It is simple to see that the CP weights  obey
the crossing symmetry relations
\be
W_{r\,s}(a) &=&\bW_{s^*\,r}(a),\quad  \bW_{r\,s}(a) =
W_{s^* \,r} (-a) \,, \label{eq:crossing} \\
\hbox{where}\quad 
 (x,y,\mu)^*&=&(\omega^{-1} y, x,1/\mu),\nonumber\ee
which can be indicated graphically by 
$$
\begin{tikzpicture}[baseline=-3pt,scale=0.6]
  \draw[aline] (-2,1) node[left] {$r$} --  (0,-1);
  \draw[aline] (0,1) node[left] {$s$} --  (-2,-1);
  \draw[bline] (-2,0) node[left] {$a$} -- (0,0) node[right] {$b$} ;
  \draw (-2,0) node[bblob] {}; \draw (0,0) node[bblob] {};
\end{tikzpicture}\quad 
=
\begin{tikzpicture}[baseline=-3pt,scale=0.6]
  \draw[aline] (-2,1) node[left] {$r$} --  (0,-1);
  \draw[aline] (-2,-1) node[left] {$s^*$} --   (0,1);
  \draw[bline] (-2,0) node[left] {$a$} -- (0,0) node[right] {$b$} ;
  \draw (-2,0) node[bblob] {}; \draw (0,0) node[bblob] {};
\end{tikzpicture},
\quad \quad
\begin{tikzpicture}[baseline=-3pt,scale=0.6]
  \draw[aline] (-2,1) node[left] {$r$} --  (0,-1);
  \draw[aline] (0,1) node[left] {$s$} --  (-2,-1);
  \draw[bline] (-1,1) node[above] {$a$} -- (-1,-1) node[below] {$b$} ;
  \draw (-1,-1) node[bblob] {}; \draw (-1,1) node[bblob] {};
\end{tikzpicture}
=
\begin{tikzpicture}[baseline=-3pt,scale=0.6]
  \draw[aline] (-2,1) node[left] {$r$} --  (0,-1);
  \draw[aline] (-2,-1) node[left] {$s^*$} -- (0,1);
  \draw[bline] (-1,-1) node[below] {$b$} -- (-1,1) node[above] {$a$} ;
  \draw (-1,-1) node[bblob] {}; \draw (-1,1) node[bblob] {};
\end{tikzpicture} \quad.
$$

\subsection{Alternative parameterisation}

The spectral parameter $\xi=(x,y,\mu)$ subject to the conditions~\eqref{eq:Ck} can be conveniently reparameterised as 
\begin{equation} \label{eq:param}
  x = e^{i(u+\phi)/N} \,,
  \qquad
  y = e^{i(u-\phi+\pi)/N} \,,
  \qquad
  \mu = e^{i(\bar\phi-\phi)/N} \,,
\end{equation}
where the variables $(\phi,\bar\phi,u)$ are now related by
\begin{equation} \label{eq:Ck2}
  \sin\phi = -k\sin u \,,
  \qquad \sin\bar\phi = -\frac{ik}{k'} \cos u \,,
  \qquad \cos\phi = k'\ \cos\bar\phi \,.
\end{equation}
Note that~\eqref{eq:Ck2} amounts to two independent relations, so that one complex parameter among $(\phi,\bar\phi,u)$ remains free.
\bigskip

When approaching the self-dual line $\{\phi=\bar\phi,k'=1 \}$, it is convenient to scale $u$ as
$$
u = -i\log k + \frac{\pi}{2} + u' \,,
$$
where $u'$ is finite. In the limit $k' \to 1$, one then gets, on the self-dual line:
$$
\sin\phi = \sin\bar\phi = -\frac{e^{-iu'}}{2} \,.
$$

\subsection{Transfer matrix and spin-chain Hamiltonian}

In this paragraph, we consider a square lattice $\cal L$ tilted by 45$^{\rm o}$, so that the rapidity lines go along the horizontal and vertical directions, unlike in \figref{CPlattice}. We denote by $\mathcal{T}_{rs}$ the transfer matrix comprising two horizontal rapidity lines with spectral parameter $\xi_r$, and $2L$ vertical rapidity lines with spectral parameter $\xi_s$, and we impose periodic boundary conditions in the horizontal direction. Then, as a consequence of the star-triangle equations~\eqref{eq:star-triangle}, the transfer matrices with different values of the horizontal parameter commute:
\begin{equation}
  {[\mathcal{T}_{r_1,s}, \mathcal{T}_{r_2,s}]} = 0 \,.
\end{equation}
Moreover, when $\xi_r=\xi_s$, the transfer matrix $\mathcal{T}_{rs}$ reduces to a cyclic translation $e^{-iP}$. Hence, one can define the associated Hamiltonian in the usual way in the limit $\xi_r \to \xi_s$, by writing:
\begin{equation} \label{eq:aniso-limit}
  \mathcal{T}_{rs} = e^{-iP} \times \left\{ \id - (u_r-u_s) \mathcal{H}_s + O[(u_r-u_s)^2] \right\} \,.
\end{equation}
This results in the spin chain Hamiltonian acting on the tensor space $V_L=\mathbb{C}^N \otimes \dots \otimes \mathbb{C}^N$ ($L$ times):
\begin{equation} \label{eq:H}
  \begin{aligned} 
    \mathcal{H}_s &= \frac{1}{N \cos \bar\phi_s}\sum_{j=1}^L \sum_{n=1}^{N-1} \left[
      \bar\alpha_n (Z_j)^n + \alpha_n (X_j^{\phantom\dag} X_{j+1}^\dag)^n
    \right] \,, \\
    \alpha_n &= \frac{\exp\left[{i(2n-N)\phi_s}/{N} \right]}{\sin({\pi n}/{N})} \,,
    \qquad
    \bar\alpha_n = k' \times \frac{\exp\left[{i(2n-N)\bar\phi_s}/{N} \right]}{\sin({\pi n}/{N})} \,.
  \end{aligned}
\end{equation}
with periodic boundary conditions $X_{L+1} \equiv X_1$. The operators $X_j$ and $Z_j$ are defined through the elementary $N \times N$ matrices $(X,Z)$ with coefficients
\begin{equation}
  X_{ab} = \w^a \ \delta_{ab} \,,
  \qquad
  Z_{ab} = \delta_{a,b-1}^{({\rm mod}\ N)} \,,
\end{equation}
which are characterised (up to a change of bases) by the relations $X^N=Z^N=\id$ and $ZX=\w XZ$.
The operator $X_j$ (resp. $Z_j$) then acts as the matrix $X$ (resp. $Z$) on the $j$-th factor of $V_L$, and as the identity matrix on the other factors. Note that \eqref{eq:H} is Hermitian for any choice of real parameters $(\phi_s,\bar\phi_s,k')$.



\subsection{$\Z_N$ charges and Kramers-Wannier duality}
\label{sec:KW}

Let $R$ be the global rotation of spins: $R= \prod_{j=1}^L Z_j$. Since $R$ and ${\mathcal H}_s$ commute, we can diagonalise them simultaneously. We denote the eigenvalues of $R$ as $\w^{-m}$ with $m \in \Z_N$, and we refer to $m$ as the $\Z_N$ charge. We may also consider the Hamiltonian~\eqref{eq:H} with ``twisted'' periodic boundary conditions $X_{L+1} \equiv \w^{\ol m} X_1$, with $\ol m \in \Z_N$, and we call $\ol m$ the dual $\Z_N$ charge.

In the present context, Kramers-Wannier (KW) duality amounts to the following non-local change of bases. 
We introduce for $j \in \{1, \dots, L\}$, the dual operators
\begin{equation}
  \ol Z_{j+1/2} = X_j^{\phantom\dag} X_{j+1}^\dag \,,
  \qquad
  \ol X_{j+1/2} = \prod_{\ell=1}^j Z_\ell^\dag \,,
\end{equation}
and we set by convention $\ol X_{1/2} = \id$. We see readily that
\begin{equation}
  \mathcal{H}_s = \sum_{j=1}^L \sum_{n=1}^{N-1} \left[
    \alpha_n (\ol Z_{j+1/2})^n + \bar\alpha_n (\ol X_{j-1/2}^{\phantom\dag} \ol X_{j+1/2}^\dag)^n
  \right] \,,
\end{equation}
with $\ol X_{j+1/2}^N = \ol Z_{j+1/2}^N = \id$, and $\ol Z_{j+1/2} \ol X_{\ell+1/2} = \w^{\delta_{j\ell}} \ol X_{\ell+1/2} \ol Z_{j+1/2}$. Hence we recover the original form~\eqref{eq:H}, except that the roles of $\alpha_n$ and $\bar\alpha_n$ have been exchanged. In terms of external parameters, KW duality acts as
\begin{equation}
  (\phi, \bar\phi, k') \longrightarrow (\bar\phi, \phi, 1/k') \,.
\end{equation}
Note that this transformation preserves the integrability condition $k'\cos\bar\phi=\cos\phi$.
Let us examine its effects on the $\Z_N$ charges $(m, \ol m)$. From the identities
\begin{equation}
  \w^{m} \ \id = \prod_{j=1}^L Z_j^\dag = \ol X_{1/2}^{\dag} \ol X_{L+1/2}^{\phantom\dag} \,,
  \qquad
  \w^{\ol m} \ \id  = X_1^\dag X_{L+1}^{\phantom\dag} = \prod_{j=1}^L \ol Z_{j+1/2}^\dag \,,
\end{equation}
it is clear that KW duality exchanges the two charges:
\begin{equation}
  (m, \ol m) \longrightarrow (\ol m, m) \,.
\end{equation}

\subsection{The Fateev-Zamolodchikov case}
\label{sec:FZ}

When $\phi=\bar\phi=0$ and $k'=1$, the chiral Potts model reduces to the Fateev-Zamolodchikov (FZ) clock model~\cite{FZclock}. In the scaling limit, the model is isotropic, and is described by the $\Z_N$-parafermionic CFT~\cite{FZcft}.

The weights of the FZ clock model enjoy the difference property, i.e., $W_{rs}(a)$ and $\bW_{rs}(a)$ are functions of $u_s-u_r$ and $a$. 
Under these conditions, the star-triangle equations~\eqref{eq:star-triangle} are consistent with the following embedding of the model in the complex plane:
\begin{equation} \label{eq:embed-graph}
  W_{rs} = \begin{tikzpicture}[baseline=-3pt,scale=0.75]
    \draw (-1,-1) circle [radius=0.08];\draw (-1,1) circle [radius=0.08];
    \draw[aline=0.9,blue] (-2,0) -- (0,0) ;
    \draw (-2,0) node[bblob] {}; \draw (0,0) node[bblob] {}; 
    \draw[dashed] (-1,-1) --  (-2,0)  -- (-1,1) -- (0,0) -- cycle;
    \draw (-1.5,-0.5) arc (-45:45:0.72cm);
    \draw (-1.9,0) node[right] {$\theta$};
  \end{tikzpicture}
  \quad, \qquad
  \bW_{rs} = \begin{tikzpicture}[baseline=-3pt,scale=0.75]
    \draw[dashed] (-1,-1) --  (-2,0) -- (-1,1) -- (0,0) -- cycle;
    \draw (-2,0) circle [radius=0.08];\draw (0,0) circle [radius=0.08];
    \draw[aline=0.8,blue] (-1,1) -- (-1,-1);
    \draw (-1,-1) node[bblob] {}; \draw (-1,1) node[bblob] {}; 
    \draw (-1.9,0) node[right] {$\theta$};\draw (-1.5,-0.5) arc (-45:45:0.72cm);
  \end{tikzpicture} \quad,
\end{equation}
where the angle $\theta$ is defined as
\begin{equation} \label{eq:embed}
  \theta = u_s-u_r \,,
\end{equation}
and the rhombi have a unit side length.
Note that this choice also respects crossing symmetry, in that the embedding angle of the left hand side of the first crossing relation \eqref{eq:crossing} is $u_s-u_r$, and the embedding angle of the right-hand-side is $\pi-u_s+u_r$ (and similarly for the second crossing relation).


\subsection{The Ising case}
\label{sec:Ising}

The chiral Potts model with $N=2$ corresponds to an Ising model with a partition function of the form
\begin{equation}
  \mathcal{Z} = \sum_{\sigma_j = \pm 1} \prod_{\aver{ij}} \exp(K_{\aver{ij}} \sigma_i \sigma_j) \,,
\end{equation}
where $K_{\aver{ij}}=K_1$ (resp. $K_{\aver{ij}}=K_2$) if $\aver{ij}$ is a horizontal (resp. vertical) edge. In this case, the algebraic relations~\eqref{eq:Ck} can be parametrised by Jacobi elliptic functions of modulus $k$ \cite{Baxter-book}:
\begin{equation} \label{eq:param-Ising}
  x = -\sqrt{k} \sn\beta \,,
  \qquad y = -\sqrt{k} \frac{\cn\beta}{\dn\beta} \,,
  \qquad \mu = \frac{\sqrt{k'}}{\dn\beta} \,.
\end{equation}
For any value of $k'$, the couplings are functions of $\beta_s-\beta_r$:
\begin{equation} \label{eq:K-Ising}
  e^{-2K_1} = k' {\rm scd}(K-\beta_s+\beta_r) \,,
  \qquad
  e^{-2K_2} = k' {\rm scd}(\beta_s-\beta_r) \,,
\end{equation}
where ${\rm scd}(u) = \sn\frac{u}{2} / (\cn\frac{u}{2} \dn\frac{u}{2})$, and $K$ is the complete elliptic integral of the first kind of modulus $k$. Using elliptic identities, one gets the relation
\begin{equation}
  \sinh 2K_1 \ \sinh 2K_2 = \frac{1}{k'} \,.
\end{equation}

Expressions~\eqref{eq:K-Ising} show that the Ising model enjoys the difference property for any value of $k$. 
Requiring that the star-triangle relation can be represented as a geometric relation in the complex plane 
leads us to define the embedding angle (see \secref{FZ}) as
\begin{equation} \label{eq:embed2}
  \theta_k = \frac{\pi}{K} (\beta_s - \beta_r) \,.
\end{equation}
From~\eqref{eq:param} and \eqref{eq:param-Ising}, we have the relations
\begin{equation} \label{eq:param-Ising2}
  e^{iu} = -\frac{i k \sn\beta \cn\beta}{\dn\beta} \,,
  \qquad e^{i\phi} = \frac{i \sn\beta \dn\beta}{\cn\beta} \,,
  \qquad e^{i\bar\phi} = \frac{ik' \sn\beta}{\cn\beta\dn\beta} \,.
\end{equation}
In the critical limit $k \to 0$, a way to recover the expression~\eqref{eq:embed} for the embedding angle is to set
\begin{equation} \label{eq:scaling-beta}
  \beta = \frac{K}{\pi} \left( \frac{i}{2} \log p + 2\beta' \right) \,,
\end{equation}
where $\beta'$ is finite [$p$ is the nome of elliptic functions in~\eqref{eq:param-Ising}, with $k \sim 4p^{1/2}$].
In this regime, we have the expansions of elliptic functions (see~\cite{Baxter-book}):
\begin{equation} \label{eq:elliptic}
  \begin{aligned}
    & H(\beta) = -i e^{i\beta'} + i e^{-i\beta'} p^{1/2} + O(p^{3/2}) \,, \\
    & H_1(\beta) = e^{i\beta'} + e^{-i\beta'} p^{1/2} + O(p^{3/2}) \,, \\
    & \Theta(\beta) = 1 - e^{2i\beta'} p^{1/2} + O(p^{3/2}) \,, \\
    & \Theta_1(\beta) = 1 + e^{2i\beta'} p^{1/2} + O(p^{3/2}) \,.
  \end{aligned}
\end{equation}
Using these, we find that $e^{iu} = -e^{2i\beta'} +O(p^{3/2})$, and hence \eqref{eq:embed2} reduces to~\eqref{eq:embed} up to corrections of order $k^3$.


\subsection{Critical behaviour}

Returning to the general $N \geq 3$ case, we shall describe the critical behaviour of the chiral Potts Hamiltonian~\eqref{eq:H} in the space of parameters $(\phi,\bar\phi,k')$.
\bigskip

On each of the self-dual (SD) lines
\begin{equation} \label{eq:SD}
  {\rm SD}_1 = \{ \phi=\bar\phi, \ k'=1 \} 
  \qquad \text{and} \qquad
  {\rm SD}_2 = \{ \phi=-\bar\phi, \ k'=1 \} \,,
\end{equation}
the model is invariant (modulo a global spin reversal) under KW duality, and thus it is massless. On each of the SD lines, for $\phi \neq 0$, the model is in an ``incommensurate state'', (i.e. the ground state is in a sector with non-zero momentum), and remains critical~\cite{Cardy93}, with the same critical exponents as in the isotropic $\Z_N$-parafermionic CFT, but its correlations become anisotropic in Minkowski space. These SD lines correspond respectively to a perturbation of the $\Z_N$-parafermionic CFT by an operator of conformal spin $+1$ and $-1$. 
\bigskip

As shown in~\cite{McCoyRoan90,Alb89a,Alb89b,Alb91}, in the plane $\phi=\bar\phi$, the model remains massless in a limited region around SD$_1$, and then it undergoes a commensurate-incommensurate transition to a massive phase. It seems plausible that these results also hold outside this plane, i.e., there is a massless incommensurate phase around the plane $\{k'=1\}$, and the model becomes massive for $|k'-1|$ large enough. At the FZ point, the massless phase is ``pinched'', and any small perturbation outside the plane $k'=1$ develops a finite mass.

\section{Non-local Operators and Quantum Groups}
\label{sec:NLQGC}

In this section, we review the picture of non-local operators arising from quantum groups that was developed by Bernard and Felder in \cite{BF91}. In \cite{IWWZ}, these currents were used in a direct way to construct discretely holomorphic operators in dense and dilute loop models. In the case of chiral Potts, we will show in Section \ref{sec:NLCCPM} that the currents constructed using the method of \cite{BF91} split naturally into two half-currents which obey a discrete holomorphicity condition.

The starting point of Bernard and Felder \cite{BF91} is to consider a quasi-triangular Hopf algebra ${\cal A}$ (also known as a quantum group) defined in terms of a set of generators $\{J_a, \thab{a}{b}, \thabhat{a}{b}\}$, $a,b=1,2,\ldots,n$ that have the relations
\be
\thab{a}{b} \thabhat{c}{b}= \delta_{a,c}
\quad
\hbox{and}
\quad
\thabhat{b}{a} \thab{b}{c}= \delta_{a,c}
\label{eq:inversion}
\ee
and the coproduct structure
\ben
\Delta(J_a) = J_a \otimes 1 + \Theta_a{}^b\otimes J_b \,,\quad 
\Delta(\Theta_a{}^b)=\Theta_a{}^c\otimes \Theta_c{}^b \,, \quad 
\Delta(\widehat\Theta^a{}_b)=\widehat\Theta^a{}_c\otimes \widehat\Theta^c{}_b \,,
\een
where $\Delta:\cA\ra \cA\ot \cA$ denotes the coproduct (for a gentle introduction to quantum groups see for example \cite{jimbo1992}). Note that we use the convention that repeated indices are summed over. The antipode and counit that complete the Hopf algebra structure can be found in \cite{BF91}.

It is helpful to introduce a graphical notation for representations of $\cA$. We indicate a representation of $\cA$ by a line and the action of the above generators of $\cA$ on this representation by
\begin{equation*}
J_a=\begin{tikzpicture}[baseline=-3pt,scale=0.75]
\draw(1,1) -- (1,-1);
\draw[wavy]
(0,0) node[below] {$a$} -- (1,0) node[oper] {} ;
\end{tikzpicture}
\quad ,
\quad\quad
\thab{a}{b}=\begin{tikzpicture}[baseline=-3pt,scale=0.75]
\draw (1,1) -- (1,-1);
\draw[wavy]
 (0,0) node[below] {$a$}-- (2,0) node[below] {$b$}  ;
\end{tikzpicture}\,,
\quad\quad
\thabhat{a}{b}=\begin{tikzpicture}[baseline=-3pt,scale=0.75]
\draw (1,1) -- (1,-1);
\draw[wavy]
 (2,0) node[below] {$b$} --(0,0) node[below] {$a$} ;
\end{tikzpicture}\quad .
\end{equation*}
Adopting the convention that composition $A\circ B$ means that $B$ is above $A$, the inversion relations \eqref{eq:inversion} are then represented as 
\begin{equation}
\begin{tikzpicture} [baseline=-3pt,scale=0.75]
\draw (0,-0.75) -- (0,0.75);
\draw[wavy=0.35](1,0) -- (1,0.3) --(-1,0.3);
\draw[wavy=0.4] (-1,-0.3) -- (1,-0.3)--(1,0);
\end{tikzpicture}
=
\begin{tikzpicture} [baseline=-3pt,scale=0.75]
\draw[wavy] (-1,-0.3) -- (-0.5,-0.3) to[out=0,in=0] (-0.5,0.3) -- (-1,0.3);
\draw (0,0.75) -- (0,-0.75);
\end{tikzpicture}
\qquad \hbox{and} \qquad
\begin{tikzpicture} [baseline=-3pt,scale=0.75,xscale=-1]
\draw (0,-0.75) -- (0,0.75);
\draw[wavy=0.35] (1,0) -- (1,0.3) -- (-1,0.3);
\draw[wavy=0.4] (-1,-0.3) -- (1,-0.3) -- (1,0);
\end{tikzpicture}
=
\begin{tikzpicture} [baseline=-3pt,scale=0.75,xscale=-1]
\draw[wavy] (-1,-0.3) -- (-0.5,-0.3) to[out=0,in=0] (-0.5,0.3) -- (-1,0.3);
\draw (0,-0.75) -- (0,0.75);
\end{tikzpicture}\quad\quad.\label{eq:unitarity}
\end{equation}
The action of the generators on tensor products of representations is indicated by
\ben
\Delta(J_a)&=&
\begin{tikzpicture}[baseline=-3pt,scale=0.75]
\draw(1,-1) node[below]{   \quad$J_a \otimes 1$} -- (1,1);\draw(2,-1) -- (2,1);
\draw[wavy]
(0,0) node[below] {$a$} -- (1,0) node[oper] {} ;
\end{tikzpicture} \quad  +
\begin{tikzpicture}[baseline=-3pt,scale=0.75]
\draw(1,-1) node[below]{\quad\quad$\Theta_a{}^b\otimes J_b$} -- (1,1);\draw(2,-1)  -- (2,1);
\draw[wavy]
(0,0) node[below] {$a$} -- (2,0) node[oper] {} ;\quad,
\end{tikzpicture}\quad,
\\
\quad
\Delta(\thab{a}{b})&=&\begin{tikzpicture}[baseline=-3pt,scale=0.75]
\draw(1,-1) node[below]{\quad\quad$\Theta_a{}^c\otimes \Theta_c{}^b$} -- (1,1);\draw(2,-1) -- (2,1);
\draw[wavy]
 (0,0) node[below] {$a$}-- (3,0) node[below] {$b$}  ;
\end{tikzpicture}\quad,
\quad\quad
\Delta(\thabhat{a}{b})=\begin{tikzpicture}[baseline=-3pt,scale=0.75]
\draw(1,-1) node[below]{\quad\quad$\widehat\Theta^a{}_c\otimes \widehat\Theta^c{}_b$} -- (1,1);\draw(2,-1) -- (2,1);
\draw[wavy]
 (3,0) node[below] {$b$} --(0,0) node[below] {$a$} ;
\end{tikzpicture}\quad.
\een
Denoting the above coproduct by $\Delta^{(2)}:\cA\ra\cA\ot\cA$ , it is then possible to define a coproduct $\Delta^{(L)}:\cA\ra \cA\ot \cA \ot \cdots \ot \cA$ (with $L$ terms in the tensor product on the right) recursively by $ \Delta^{(m+1)}=(\Delta \ot \id \ot \id \ot \cdots \cdots \id)  \Delta^{(m)}$. Then it follows that the representation of $\Delta^{(L)}(J_a)$ on an $L$-fold tensor product is indicated graphically by
\ben\Delta^{(L)}(J_a)=\sum\limits_{i=1}^L
\begin{tikzpicture}[baseline=-3pt,scale=0.75]
 \draw(0,-1) -- (0,1);
\draw(1,-1) -- (1,1);
\draw(2,-1)  -- (2,1);
\draw(3,-1)  -- (3,1);
\draw(4,-1)  -- (4,1);
\draw(5,-1)  -- (5,1);
\draw[wavy] (-1,0) node[below] {$a$} -- (2,0) node[oper] {} ;
\draw(2,1) node[above] {$i$};
\end{tikzpicture}
\een

The above graphics becomes more useful when combined with the standard graphics for the R-matrix. 
The R-matrix of a quantum group, $\check{R}:V_1\ot V_2\rightarrow V_2\ot V_1$, is a map between tensor products of representations
$V_1$ and $V_2$ that commutes with the action of the quantum group $\cA$. That is, we have
$ \check{R}\, \Delta(x)=\Delta(x)\, \check{R}$. We can represent the R-matrix graphically by
\ben
\begin{tikzpicture}[baseline=-3pt,scale=0.65]
\draw[arr=0.25] (-1,1) node[left] {$1$} -- (1,-1);
\draw[arr=0.25] (1,1) node[right] {$2$} -- (-1,-1);  
\end{tikzpicture},\een
where the arrows serve to orient the picture and we view the R-matrix above as acting from top to bottom.
We shall generally suppress these arrows. The commutation relations $ \check{R}\, \Delta(x)=\Delta(x) \,\check{R}$ then have
the following simple graphical realisations when  $x=J_a,\thab{a}{b},$ and $\thabhat{a}{b}$:
\vspace*{3mm}

\be
\hspace*{-15mm}  \begin{tikzpicture} [baseline=-3pt,scale=0.65]
    \draw (-1,1) -- (1,-1);
     \draw (1,1) -- (-1,-1);
    \draw (0,-2) node[below]{$\quad \check{R} (J_a \otimes 1)\quad+\quad$};
    \draw[wavy=0.3] (-2,0.5) node[left] {$a$} -- (-0.5,0.5) node[oper] {};
  \end{tikzpicture}\hspace*{-8mm}
  +
  \begin{tikzpicture} [baseline=-3pt,scale=0.65]
    \draw (-1,1) -- (1,-1);
     \draw (1,1) -- (-1,-1);
    \draw (0,-2) node[below]{$\check{R} (\thab{a}{b}\otimes J_b)\quad\quad\quad=$};
    \draw[wavy=0.3] (-2,0.5) node[left] {$a$} -- (0.5,0.5) node[oper] {};
  \end{tikzpicture}
  =
 \begin{tikzpicture} [baseline=-3pt,scale=0.65]
    \draw (-1,1) -- (1,-1);
     \draw (1,1) -- (-1,-1);
    \draw (0,-2) node[below]{$( J_a\otimes 1) \check{R}\quad\quad\quad+$};
    \draw[wavy=0.3] (-2,-0.5) node[left] {$a$} -- (-0.5,-0.5) node[oper] {};
  \end{tikzpicture}\hspace*{-8mm}
  +
  \begin{tikzpicture} [baseline=-3pt,scale=0.65]
    \draw (-1,1) -- (1,-1);
     \draw (1,1) -- (-1,-1);
    \draw (0,-2) node[below]{$(\thab{a}{b}\otimes J_b) \check{R},$};
    \draw[wavy=0.3] (-2,-0.5) node[left] {$a$} -- (0.5,-0.5) node[oper] {};
  \end{tikzpicture}
\label{eq:Jcom}
\ee
and
\be
\hspace*{-15mm} \begin{tikzpicture} [baseline=-3pt,scale=0.65]
  \draw (-1,1) -- (1,-1);
  \draw (1,1) -- (-1,-1);
  \draw[wavy=0.2] (-1.5,0.5) node[left] {$a$} -- (1.5,0.5) node[right] {$b$};
  \draw (0,-2) node[below]{$\quad \check{R} (\thab{a}{c} \otimes \thab{c}{b})\quad\quad=$};
\end{tikzpicture}\hspace*{-5mm}
=
\begin{tikzpicture} [baseline=-3pt,scale=0.65]
  \draw (-1,1) -- (1,-1);
  \draw (1,1) -- (-1,-1);
  \draw[wavy=0.2] (-1.5,-0.5) node[left] {$a$} -- (1.5,-0.5) node[right] {$b$};
  \draw (0,-2) node[below]{$ (\thab{a}{c} \otimes \thab{c}{b}) \check{R}\quad\quad$};
\end{tikzpicture}, \quad
\begin{tikzpicture} [baseline=-3pt,scale=0.65]
  \draw (-1,1) -- (1,-1);
  \draw (1,1) -- (-1,-1);
  \draw[wavy=0.2] (1.5,0.5) node[right] {$b$} -- (-1.5,0.5) node[left] {$a$};
  \draw (0,-2) node[below]{$\check{R} (\thabhat{a}{c} \otimes \thabhat{c}{b})=$};
\end{tikzpicture}\hspace*{-5mm}
=
\begin{tikzpicture} [baseline=-3pt,scale=0.65]
  \draw (-1,1) -- (1,-1);
  \draw (1,1) -- (-1,-1);
  \draw[wavy=0.2] (1.5,-0.5) node[right] {$b$} -- (-1.5,-0.5) node[left] {$a$};
  \draw (0,-2) node[below]{$(\thabhat{a}{c} \otimes \thabhat{c}{b}) \check{R}.$};
\end{tikzpicture}
\label{eq:thetacom}
\ee

We now wish to define a non-local operator $j_a(x,y)$ associated with the insertion of the operator $J_a$ at
a point $(x,y)$ in a 2D lattice model and an attached tail made up of tensor products of $\thab{a}{b}$ and $\thabhat{a}{b}$ along some path leading to a marked point on the boundary of the lattice. In order to do this, we again follow the approach of Bernard and Felder. Suppose we have a 2D lattice $\Lambda$ consisting of 4-vertices at points $\vec p\in \R^2$. Let $\Lambda'$ denote the lattice consisting of the points $\vec r$ which are the midpoints of the edges of $\Lambda$. Then if $V(\vec r)$ denotes a $\cA$ representation associated with the midpoint $\vec r$, the R-matrix associated with the vertex $\vec p\in \Lambda$ will be a map
\be \check{R}(\vec p): V(\vec r_1)\ot V(\vec r_4) \ra V(\vec r_2) \ot V(\vec r_3) \quad
\quad \begin{tikzpicture}[baseline=-3pt,scale=0.65]
\draw[arr=0.25] (-1,1) node[left] {$\vec r_1$} -- (1,-1) node[right] {$\vec r_3$} ;
\draw[arr=0.25] (1,1) node[right] {$\vec r_4$} -- (-1,-1) node[left] {$\vec r_2$};
\draw (-0.25,0) node[left] {$\vec p$};  
\end{tikzpicture},\label{eq:Rmatrix2}\ee 
where the $\vec r_i\in \Lambda'$ are the indicated four midpoints surrounding the point $\vec p$  (some of the $V_{\vec r_i}$ will need to be isomorphic for the R-matrix to exist, which we assume to be the case). We can then define a vector space  
$$
V_{\Lambda} = \bigotimes\limits_{\vec r\in \Lambda'} V(\vec r) \,,
$$
and a linear operator $B: V_{\Lambda}\ra V_{\Lambda}$, by
$$
B=\bigotimes\limits_{\vec p\in \Lambda} \check{R}(\vec p) \,.
$$
The partition function of the vertex model with Boltzmann weights specified by the R-matrices is given by 
\be
\mathcal{Z} = \hbox{Tr}_{V_\Lambda}(B) \,,
\label{eq:partfn}
\ee
and the expectation value of any linear operator ${\cal{O}}:V_\Lambda \ra V_\Lambda$ is given by
\be
\langle{\cal{O}}\rangle = \frac{1}{\mathcal{Z}} \hbox{Tr}_{V_\Lambda} ({\cal{O}} B) \,.
\label{eq:expvalue}
\ee

The above expressions \eqref{eq:partfn} and \eqref{eq:expvalue}  may at first seem unusual -- the partition function and correlation functions are more commonly given as the trace over a 1D tensor product space that would correspond to the Hilbert space in the quantum statistical mechanics interpretation of the partition function. However, it is a useful, and simple, exercise to check that the partition function does always reduce to the standard 1D trace. In contrast, the expectation value $\langle {\cal{O}}\rangle$ may  be written as a 1D trace  only in the special case when the operator ${\cal{O}}$ acts trivially at all $\vec r$ in $V_\Lambda$ except those along some 1D line in the lattice. 

\begin{figure}
  \label{fig:pathins}
  \centering\hspace*{-10mm}
  \begin{tikzpicture} [baseline=-3pt,scale=0.3]
    \draw (0,0) -- (16,16);
    \draw (4,0) -- (16,12);
    \draw (8,0) -- (16,8);
    \draw (12,0) -- (16,4);
    \draw (0,4) -- (12,16);
    \draw (0,8) -- (8,16);
    \draw (0,12) -- (4,16);
    \draw (0,2) -- (2,0);
    \draw (0,6) -- (6,0);
    \draw (0,10) -- (10,0);
    \draw (0,14) -- (14,0);
    \draw (2,16) -- (16,2);
    \draw (6,16) -- (16,6);
    \draw (10,16) -- (16,10);
    \draw (14,16) -- (16,14);
    \draw[wavy=0.52] (0,11) node[left] {$a$} -- (1,11) -- (5,7) -- (3,5) -- (5,3) -- (9,7) --
    (5,11) -- (7,13) -- (12,8) node[oper] {}; 
    \draw (7,5) node[below] {$\;\;\gamma$};
    \draw (12,8) node[below] {$\; \;\;\vec r$};

  \end{tikzpicture}
  \caption{The insertion points and path of a non-local operator $j_a^{\gamma}(\vec r)$.}
\end{figure}
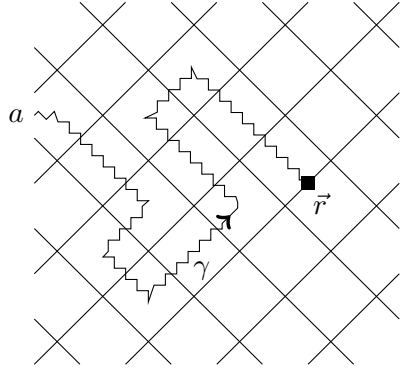

In this paper, we are interested in operators $\cO$ that act non-trivially on a set of midpoints $\vec r$ that will wind along a path $\gamma$ from a marked interior point to a marked point on the boundary of the lattice. In this case the more general expression \eqref{eq:expvalue} will be required. To be more precise, we consider the operator $j_a^\gamma(\vec r):V_\Lambda\ra V_\Lambda$ constructed by the insertion of the representation of $J_a$ on $V(\vec r)$ at the midpoint $\vec r$, and the insertion of a ``tail operator'' constructed from the insertion of $\thab{a}{b}$ or $\thabhat{a}{b}$ along a line of midpoints specified by the path $\gamma$ that terminates at some fixed, but arbitrary, point on the boundary of the lattice. An example is shown in \figref{pathins}.

The commutation relations with the R-matrix expressed by \eqref{eq:Jcom} and \eqref{eq:thetacom} have two immediate consequences for expectation values $\langle j_a^\gamma(\vec r)\rangle$.
The second relation \eqref{eq:thetacom} implies that the expectation value is independent of the path $\gamma$ and will depend only upon the insertion point $\vec r$ and the fixed boundary point. Thus we will from now on drop the $\gamma$ path superscript on the current. The other commutation relation \eqref{eq:Jcom}  implies that when inserted into an expectation value we have
\begin{equation} \label{eq:jdh}
  j_a(\vec r_1) - j_a(\vec r_2) - j_a(\vec r_3) + j_a(\vec r_4)=0 \,,
\end{equation}
where $\vec r_i$ are the four edge midpoints points surrounding any vertex -- as indicated in \eqref{eq:Rmatrix2}. After embedding the lattice into the complex plane it is this relation \eqref{eq:jdh} that was
interpreted as a discrete holomorphicity relation in several examples in the paper \cite{IWWZ}.

\section{Chiral Potts Weights and Representation Theory}
\label{sec:CPRT}

In this section we review the construction of chiral Potts weights in terms of the 
representation theory of $\uq$ \cite{DJMM90sl3,date1990new,DJMM91gln}. The notation we use is that
of \cite{date1990new}.

\subsection{The quantum affine algebra $\uqt$}

We begin by defining $\uqt$, which is an algebra over $\mathbb{C}$ generated by $e_i,f_i,t_i^{\pm 1},z_i$ $(i=0,1)$, where $e_i,f_i,t_i^{\pm 1}$ satisfy the standard relation of the quantum affine algebra $\uq$, and $z_0$ and $z_1$ are two new central elements. The comultiplication of $\uqt$ is chosen as 
\ben \Delta(e_i)&=&e_i\ot \id + z_i t_i\ot e_i,\quad \Delta(f_i)=f_i\ot t_i^{-1} + z_i^{-1}\ot f_i,\\
\Delta(t_i)&=&t_i \ot t_i,\quad \Delta(z_i)=z_i\ot z_i.
\een
The representations relevant to the chiral Potts occur when $q=-\exp(i\pi/N)$ where $N\in \{2,3,4,\cdots\}$, and
we shall fix $q$ to take this value from now on. We also define $\w=q^2$. These representations are $N$-dimensional cyclic representations denoted $V_{rr'}$ and parametrised by a pair of points $(r,r')\in \mathcal{C}_k \times \mathcal{C}_k$.
Here $\cC_k$ is the algebraic curve~\eqref{eq:Ck} given by $(x,y,z)\in \C^3$ such that
\ben
x^N+y^N=k(1+x^N y^N) \,,
\quad
\mu^N=\frac{k'}{1-k x^N}=\frac{1-ky^N}{k'} \,,\;\;
\een
where $k^2+k'^2 = 1$.

If $r=(x,y,z)\in \cC_k$ and $r'=(x',y',z')\in \cC_k$ then the representation $V_{rr'}$ is given by
\begin{equation} \label{eq:cyclrep}
  \begin{aligned}
    & \pi_{rr'} (e_1)= \frac{q }{(q^2-1)^2} (x\mu\mu'Z-y')X \,,
    \qquad \pi_{rr'} (f_1)= \frac{c_0}{x x' \mu \mu'} X^{-1} (yZ^{-1}-x' \mu \mu') \,,  \\
    & \pi_{rr'} (t_1)= c_0\mu\mu' Z,\squad \pi_{rr'} (z_1)=c_0^{-1}, \\
    & \pi_{rr'} (e_0)= \frac{q}{(q^2-1)^2} X^{-1} (y(\mu \mu')^{-1}Z^{-1}-x') \,,
    \quad \pi_{rr'} (f_0) = \left(\frac{c_0 \mu\mu'}{x'}Z - \frac{q^2}{c_0 y}\right)X \,, \\
    & \pi_{rr'} (t_0)= \frac{1}{c_0 \mu\mu'} Z^{-1},\squad \pi_{rr'} (z_0)=c_0 \,.
  \end{aligned}
\end{equation}
Here, the  objects $X$ and $Z$ are $N\times N$ matrices, such that $ZX=\omega X Z$ and $X^N=Z^N=\id$. In this paper, we shall fix  $X$ and $Z$ as
\ben 
X=\begin{pmatrix}
  1&0&0&\cdots&0&0\\
  0&\w&0&\cdots &0&0\\
  0&0&\w^2&\cdots &0&0\\
  \vdots&\vdots&\vdots&\vdots &\vdots&\vdots\\
  0&0&0&\cdots &0&\w^{N-1}\\
\end{pmatrix},
\qquad Z=\begin{pmatrix} 
  0&1&0&\cdots &0&0\\
  0&0&1&0&\cdots &0\\
  \vdots&\vdots&\vdots&\vdots &\vdots&\vdots\\
  0&0&0&\cdots &0&1\\
  1&0&0&\cdots &0&0\\
\end{pmatrix}
\een
The constant\footnote{The other constants appearing in Section 4 of \cite{date1990new} are here fixed as $c_1=1/c_0$ and $\kappa_i=1/(q^2-1)$} $c_0$ also appearing in the above satisfies $c^2_0={q^2 x x'}/{(y y')}$.

\subsection{The $\check{R}$-matrix}

Consider now the R-matrix  $\check{R}(rr',ss'): V_{rr'}\ot  V_{ss'}\rightarrow  V_{ss'} \ot  V_{rr'}$ that obeys $\check{R}(rr',ss') \Delta(x)=\Delta(x) \check{R}(rr',ss')$ for $x \in \uqt$. The approach of \cite{date1990new} starts with the Ansatz that this R-matrix is of the factorised form
\be
\check{R}(rr',ss')= S_{r's}(T_{r's'}\ot T_{rs}) S_{rs'} \,, \label{eq:factR}
\ee
where 
\ben
S_{rs'}:  V_{rr'}\ot  V_{ss'} \rightarrow V_{s'r'}\ot  V_{sr} \,,
\quad T_{rs}: V_{sr}\rightarrow V_{rs} \,.
\een
Then the relation
$$
\check{R}(rr',ss')[\pi_{rr'}\ot \pi_{ss'}(\Delta(x))]
=[\pi_{ss'}\ot \pi_{rr'}(\Delta(x))] \check{R}(rr',ss')
$$
is ensured if $S$ and $T$ satisfy the stronger `sufficiency conditions':
\be 
\hspace*{-10mm} S_{rs'}[\pi_{rr'}\ot \pi_{ss'}(\Delta(x))]
&=& [\pi_{s'r'}\ot \pi_{sr}(\Delta(x))] S_{rs'} \,, \label{eq:suffcond1} \\
\hspace*{-10mm} ( T_{r's'}\ot 1)   [\pi_{s'r'}\ot \pi_{sr}(\Delta(x))]
&=& [\pi_{r's'}\ot \pi_{sr}(\Delta(x))]( T_{r's'}\ot 1) \,,  \label{eq:suffcond2} \\
\hspace*{-10mm} (1\ot T_{rs})  [\pi_{r's'}\ot \pi_{sr}(\Delta(x))]
&=&  [\pi_{r's'}\ot \pi_{rs}(\Delta(x))](1\ot T_{rs}) \,,  \label{eq:suffcond3} \\
\hspace*{-10mm} S_{r's} [\pi_{r's'}\ot \pi_{rs}(\Delta(x))]
&=&  [\pi_{ss'}\ot \pi_{rr'}(\Delta(x))] S_{r's} \,.  \label{eq:suffcond4}
\ee
These sufficiency conditions were in turn found to be satisfied by the choice 
$$
\hspace*{-10mm}
S_{rs}(v_{\ep_1}\ot v_{\ep_2}) = W_{rs}(\ep_1-\ep_2) (v_{\ep_2}\ot v_{\ep_1}) \,,
\qquad T_{rs} v_\ep= \sli_{a=0}^{N-1} \bW_{rs}(a) v_{\ep-a} \,,
$$
with the coefficients given by
\ben 
\frac{W_{rs}(n)}{W_{rs}(n-1)}=
\frac{\mu_r}{\mu_s}\frac{y_s-x_r \w^{n}}{y_r-x_s \w^{n}} \,,
\qquad
\frac{\bW_{rs}(n)}{\bW_{rs}(n-1)}=
\mu_r\mu_s\frac{x_r \w-x_s \w^{n}}{y_s-y_r \w^{n}} \,,
\een
where we have now switched to the notation $a=(x_a,y_a,\mu_a)$ for a point $a\in \cC_k$. These are the precisely the chiral Potts model Boltzmann weights~\eqref{eq:W} which can be found in \cite{star-triangle88}. Defining components $R(rr',ss')^{ab}_{cd}$ of the R-matrix by $\check{R}(rr',ss')(v_a\ot v_b)=\sli_{c,d} R(rr',ss')^{ab}_{cd}(v_d\ot v_c)$, then leads via \eqref{eq:factR} to the factorised expression
\be
R(rr',ss')^{ab}_{cd}= W_{r's}(d-c) \bW_{r's'}(a-d)\bW_{rs}(b-c) W_{rs'}(a-b) \,.
\label{eq:Rfact}
\ee

It is useful to introduce a version of the standard 4-vertex graphical notation for R-matrices that is modified to deal with the representation $V_{rr'}$.
We indicate the identity acting on the representation $V_{rr'}$ by a directed double line
\begin{tikzpicture}[baseline=10pt,scale=0.5]
\draw[aline] (0,1.5) node[above] {$r'$} -- (0,0);
\draw[aline] (1,1.5) node[above] {$r$} -- (1,0);
\end{tikzpicture}\; ,\; 
and the R-matrix  $\check{R}(rr',ss')$ by 
\\[-5mm]\ben\check{R}(rr',ss') = 
\begin{tikzpicture}[baseline=-3pt,scale=0.6]
\draw[aline] (2,1) node[right] {$s$} --  (-1,-2);
\draw[aline]  (1,2) node[right] {$s'$} -- (-2,-1);
\draw[aline]  (-1,2) node[left] {$r$}--(2,-1);
\draw[aline]  (-2,1) node[left] {$r'$} --(1,-2);
\end{tikzpicture},\quad 
\een
The components  $R(rr',ss')^{ab}_{cd}$ are then indicated by
\\[-5mm]\ben R(rr',ss')^{ab}_{cd} = 
\begin{tikzpicture}[baseline=-3pt,scale=0.6]
\draw[aline] (2,1) node[right] {$s$} --  (-1,-2);
\draw[aline]  (1,2) node[right] {$s'$} -- (-2,-1);
\draw[aline]  (-1,2) node[left] {$r$}--(2,-1);
\draw[aline]  (-2,1) node[left] {$r'$} --(1,-2);
\draw[bline] (-1,-1) node[below] {$d$}-- (1,-1) node[below] {$c$};
\draw[bline] (-1,1) node[bblob] {} -- (-1,-1) node[bblob] {};
\draw[bline] (-1,1) node[above] {$a$} -- (1,1) node[above] {$b$} ;
\draw[bline] (1,1) node[bblob] {} -- (1,-1) node[bblob] {};
\end{tikzpicture},\quad 
\een
with the graphical conventions of \secref{CP}.

\section{Construction of non-local operators in the Chiral Potts model}
\label{sec:NLCCPM}

In this section, we consider the non-local operators $j_a(\vec r)$ discussed in \secref{NLQGC} associated with the cyclic representations of $\uqt$ introduced in \secref{CPRT}.

\subsection{The current $j_{\bar{e}_0}$}

Let us first consider the current associated with the generator $\bar{e}_0:=t_0f_0$. We have the co-products
\be
\Delta(\bar{e}_0)= \bar{e}_0 \otimes \id + t_0 z_0^{-1} \otimes \bar{e}_0 \,,
\quad\hbox{and}\quad \Delta(t_0 z_0^{-1})= t_0 z_0^{-1}\ot t_0 z_0^{-1} \,,
\label{eq:coprod2}
\ee
and thus $\bar{e}_0$ and  $t_0 z_0^{-1}$ are respectively generators of type $J_a$ and $\Theta_a^{\:b}$ in the 
notation of \secref{NLQGC}. It follows from \eqref{eq:cyclrep} that the action of these generators on the representation $V_{rr'}$ is given by
\be \hspace*{-10mm}
\pi_{rr'}(\bar{e}_0)&=& 
X\left[x_{r'}^{-1}-y_r^{-1}\pi_{rr'}(t_0z_0^{-1}) \right] \,, \quad
\pi_{rr'}(t_0z_0^{-1}) = f_r f_{r'}  Z^{-1} \,, \label{eq:uqaction}\\[2mm]
\hbox{where}\quad f_r&:=& \frac{y_r}{-q x_r \mu_r} \,. \nn
\ee

We now wish to follow the approach of \secref{NLQGC} and consider the non-local operator
$\bar{e}_0(x,t)$ associated with the insertion of the appropriate representation of the non-local 
operator
\be \label{eq:eobarcurrent}
\cdots \ot t_0z_0^{-1} \ot t_0z_0^{-1} \ot \bar{e}_0 \,.
\ee
The position $(x,t)$ will correspond to a CP site $(x,t)$ (indicated by a $\bullet$ in \figref{CPlattice})
and we define $\bar{e}_0(x,t)$ such that the  $\bar{e}_0$ in \eqref{eq:eobarcurrent} acts on the representation associated with a pair of diagonal lines either side of the point $(x,t)$. To make this definition more precise it is useful to modify the graphical notation of \secref{NLQGC}. To this end we introduce the following representation of the diagonal action of $X$ and non-diagonal action of $\pi_{rr'}(t_0z_0^{-1})$ on the representation $V_{rr}$:
\ben
X\quad \sim \quad  
\begin{tikzpicture}[baseline=-3pt,scale=0.6]
\draw[aline] (-0.5,1) node[left] {$r'$} --  (-0.5,-1);
\draw[aline] (0.5,1) node[left] {$r$} --  (0.5,-1);
\draw (0,0) node[boper] {};
\draw[wavy=0.5] (-2,0)--(-1,0);
\end{tikzpicture}\quad,\quad
 \pi_{(rr')}(t_0z_0^{-1}) \quad \sim  
\begin{tikzpicture}[baseline=-3pt,scale=0.6]
\draw[aline] (-0.5,1) node[left] {$r'$} --  (-0.5,-1);
\draw[aline] (0.5,1) node[left] {$r$} --  (0.5,-1);
\draw[wavy=0.5] (-1,0)--(1,0);
\end{tikzpicture}.
\een
It follows from \eqref{eq:coprod2} and \eqref{eq:uqaction} that the action of $\bar{e}_0(x,t)$ splits into two
`half-currents' which can be represented graphically as
\be \bar{e}_0(x,t) = \begin{tikzpicture}[baseline=-3pt,scale=0.5]
\draw[aline] (-0.5,1) node[left] {$r'$} --  (-0.5,-1);
\draw[aline] (0.5,1) node[right] {$r$} --  (0.5,-1);
\draw (0,0) node[boper] {};
\draw[wavy=0.5] (-3,0)--(-1,0);
\draw (-4,0) node {$\cdots$};
\draw (-6,0) node {$x_{r'}^{-1}$};
\end{tikzpicture}
\begin{tikzpicture}[baseline=-3pt,scale=0.5]
\draw[aline] (-0.5,1) node[left] {$r'$} --  (-0.5,-1);
\draw[aline] (0.5,1) node[right] {$r$} --  (0.5,-1);
\draw[wavy=0.5] (-3,0)--(-1,0) -- (0,1) -- (1,0);
\draw (0,0) node[boper] {};
\draw (-4,0) node {$\cdots$};
\draw (-6,0) node {$-y_r^{-1}$};
\end{tikzpicture}
\label{eq:be0graph}\ee
Here $(x,t)$ is the CP site marked by a $\blacksquare$ and the left tail will wind through the rest of the
CP lattice. Up to considerations of boundary conditions, which we will not need in this paper, the position
of this tail is arbitrary due to the commutation of $t_0 z_0^{-1}\ot t_0 z_0^{-1} $ with the R-matrix.

Let us now consider the  sufficiency condition \eqref{eq:suffcond1} in the case when $x=\bar{e}_0$, $r'=r$ and $s'=s$. Equation \eqref{eq:suffcond1} becomes
\ben
S_{rs}[\pi_{rr}\ot \pi_{ss}(\Delta(\bar{e}_0))]
&=&
[\pi_{sr}\ot \pi_{sr}(\Delta(\bar{e}_0))] S_{rs} \,,
\een
which can be represented graphically as 
\ben
&&\hspace*{-15mm}x_r^{-1}\,
\begin{tikzpicture}[baseline=-3pt,scale=0.4]
\draw[aline] (2,1) node[right] {$s$} --  (-1,-2);
\draw[aline]  (1,2) node[above] {$s$} -- (-2,-1);
\draw[aline]  (-1,2) node[above] {$r$}--(2,-1);
\draw[aline]  (-2,1) node[left] {$r$} --(1,-2);
\draw (0,0) node[blob] {};\draw (-2,0) node[blob] {};\draw (2,0) node[blob] {} ;
\draw (0,2) node[blob] {};\draw (0,-2) node[blob] {};
\draw (2,2) node[blob] {};\draw (2,-2) node[blob] {};
\draw (-2,2) node[blob] {};\draw (-2,-2) node[blob] {} ;
\draw[wavy=0.5](-4,0)--(-2,0);
\draw  (-1,1) node[boper] {};
\end{tikzpicture}
-y_r^{-1}\,
\begin{tikzpicture}[baseline=-3pt,scale=0.4]
\draw (2,1)  --  (-1,-2);
\draw  (1,2) -- (-2,-1);
\draw  (-1,2)--(2,-1);
\draw  (-2,1) --(1,-2);
\draw (0,0) node[blob] {};\draw (-2,0) node[blob] {};\draw (2,0) node[blob] {} ;
\draw (0,2) node[blob] {};\draw (0,-2) node[blob] {};
\draw (2,2) node[blob] {};\draw (2,-2) node[blob] {};
\draw (-2,2) node[blob] {};\draw (-2,-2) node[blob] {} ;
\draw[wavy=0.5](-4,0)--(-2,0) -- (-2,2) -- (0,2);
\draw  (-1,1) node[boper] {};
\end{tikzpicture}
+x_s^{-1}\,
\begin{tikzpicture}[baseline=-3pt,scale=0.4]
\draw (2,1)  --  (-1,-2);
\draw  (1,2) -- (-2,-1);
\draw  (-1,2)--(2,-1);
\draw  (-2,1) --(1,-2);
\draw (0,0) node[blob] {};\draw (-2,0) node[blob] {};\draw (2,0) node[blob] {} ;
\draw (0,2) node[blob] {};\draw (0,-2) node[blob] {};
\draw (2,2) node[blob] {};\draw (2,-2) node[blob] {};
\draw (-2,2) node[blob] {};\draw (-2,-2) node[blob] {} ;
\draw[wavy=0.5](-4,0)--(-2,0)-- (-2,2) -- (0,2);
\draw  (1,1) node[boper] {};
\end{tikzpicture}
-y_s^{-1}\,
\begin{tikzpicture}[baseline=-3pt,scale=0.4]
\draw (2,1)  --  (-1,-2);
\draw  (1,2) -- (-2,-1);
\draw  (-1,2)--(2,-1);
\draw  (-2,1) --(1,-2);
\draw (0,0) node[blob] {};\draw (-2,0) node[blob] {};\draw (2,0) node[blob] {} ;
\draw (0,2) node[blob] {};\draw (0,-2) node[blob] {};
\draw (2,2) node[blob] {};\draw (2,-2) node[blob] {};
\draw (-2,2) node[blob] {};\draw (-2,-2) node[blob] {} ;
\draw[wavy=0.5](-4,0)--(-2,0)-- (-2,2) -- (0,2) -- (2,2) -- (2,0);
\draw  (1,1) node[boper] {};
\end{tikzpicture}
\\[3mm]
&&\hspace*{-15mm}=
x_r^{-1}\,
\begin{tikzpicture}[baseline=-3pt,scale=0.4]
\draw (2,1)  --  (-1,-2);
\draw  (1,2)   -- (-2,-1);
\draw  (-1,2) --(2,-1);
\draw  (-2,1)  --(1,-2);
\draw (0,0) node[blob] {};\draw (-2,0) node[blob] {};\draw (2,0) node[blob] {} ;
\draw (0,2) node[blob] {};\draw (0,-2) node[blob] {};
\draw (2,2) node[blob] {};\draw (2,-2) node[blob] {};
\draw (-2,2) node[blob] {};\draw (-2,-2) node[blob] {} ;
\draw[wavy=0.5](-4,0)--(-2,0);
\draw  (-1,1) node[boper] {};
\end{tikzpicture}
-y_s^{-1}\,
\begin{tikzpicture}[baseline=-3pt,scale=0.4]
\draw (2,1)  --  (-1,-2);
\draw  (1,2) -- (-2,-1);
\draw  (-1,2)--(2,-1);
\draw  (-2,1) --(1,-2);
\draw (0,0) node[blob] {};\draw (-2,0) node[blob] {};\draw (2,0) node[blob] {} ;
\draw (0,2) node[blob] {};\draw (0,-2) node[blob] {};
\draw (2,2) node[blob] {};\draw (2,-2) node[blob] {};
\draw (-2,2) node[blob] {};\draw (-2,-2) node[blob] {} ;
\draw[wavy=0.5](-4,0)--(-2,0) -- (-2,2) -- (0,2) -- (0,0);
\draw  (-1,1) node[boper] {};
\end{tikzpicture}
+x_r^{-1}\,
\begin{tikzpicture}[baseline=-3pt,scale=0.4]
\draw (2,1)  --  (-1,-2);
\draw  (1,2) -- (-2,-1);
\draw  (-1,2)--(2,-1);
\draw  (-2,1) --(1,-2);
\draw (0,0) node[blob] {};\draw (-2,0) node[blob] {};\draw (2,0) node[blob] {} ;
\draw (0,2) node[blob] {};\draw (0,-2) node[blob] {};
\draw (2,2) node[blob] {};\draw (2,-2) node[blob] {};
\draw (-2,2) node[blob] {};\draw (-2,-2) node[blob] {} ;
\draw[wavy=0.5](-4,0)--(0,0);
\draw  (1,1) node[boper] {};
\end{tikzpicture}
-y_s^{-1}\,
\begin{tikzpicture}[baseline=-3pt,scale=0.4]
\draw (2,1)  --  (-1,-2);
\draw  (1,2) -- (-2,-1);
\draw  (-1,2)--(2,-1);
\draw  (-2,1) --(1,-2);
\draw (0,0) node[blob] {};\draw (-2,0) node[blob] {};\draw (2,0) node[blob] {} ;
\draw (0,2) node[blob] {};\draw (0,-2) node[blob] {};
\draw (2,2) node[blob] {};\draw (2,-2) node[blob] {};
\draw (-2,2) node[blob] {};\draw (-2,-2) node[blob] {} ;
\draw[wavy=0.5](-4,0)--(0,0) -- (0,2) -- (2,2) -- (2,0);
\draw  (1,1) node[boper] {};
\end{tikzpicture}
\een 
Several points are worth noting about this commutation relationship:
\begin{enumerate}
\item The tail operator is naturally associated with the edges of the dual CP lattice
with vertices indicated by $\circ$.
\item We have appended a left horizontal tail to indicate that this 
relation may be embedded in the larger CP lattice with such a left tail.
\item There is cancellation of four of the terms.
\end{enumerate}
After cancellation, we arrive at the four-term relation
\bec
&&\hspace*{-15mm}
-y_r^{-1}\,
\begin{tikzpicture}[baseline=-3pt,scale=0.4]
\draw (2,1)  --  (-1,-2);
\draw  (1,2) -- (-2,-1);
\draw  (-1,2)--(2,-1);
\draw  (-2,1) --(1,-2);
\draw (0,0) node[blob] {};\draw (-2,0) node[blob] {};\draw (2,0) node[blob] {} ;
\draw (0,2) node[blob] {};\draw (0,-2) node[blob] {};
\draw (2,2) node[blob] {};\draw (2,-2) node[blob] {};
\draw (-2,2) node[blob] {};\draw (-2,-2) node[blob] {} ;
\draw[wavy=0.5](-4,0)--(-2,0) -- (-2,2) -- (0,2);
\draw  (-1,1) node[boper] {};
\end{tikzpicture}
\;+x_s^{-1}\,
\begin{tikzpicture}[baseline=-3pt,scale=0.4]
\draw (2,1)  --  (-1,-2);
\draw  (1,2) -- (-2,-1);
\draw  (-1,2)--(2,-1);
\draw  (-2,1) --(1,-2);
\draw (0,0) node[blob] {};\draw (-2,0) node[blob] {};\draw (2,0) node[blob] {} ;
\draw (0,2) node[blob] {};\draw (0,-2) node[blob] {};
\draw (2,2) node[blob] {};\draw (2,-2) node[blob] {};
\draw (-2,2) node[blob] {};\draw (-2,-2) node[blob] {} ;
\draw[wavy=0.5](-4,0)--(-2,0)-- (-2,2) -- (0,2);
\draw  (1,1) node[boper] {};
\end{tikzpicture}
\\[3mm]
&&\hspace*{-15mm}=
-q^2 y_s^{-1}\,
\begin{tikzpicture}[baseline=-3pt,scale=0.4]
\draw (2,1)  --  (-1,-2);
\draw  (1,2) -- (-2,-1);
\draw  (-1,2)--(2,-1);
\draw  (-2,1) --(1,-2);
\draw (0,0) node[blob] {};\draw (-2,0) node[blob] {};\draw (2,0) node[blob] {} ;
\draw (0,2) node[blob] {};\draw (0,-2) node[blob] {};
\draw (2,2) node[blob] {};\draw (2,-2) node[blob] {};
\draw (-2,2) node[blob] {};\draw (-2,-2) node[blob] {} ;
\draw[wavy=0.5](-4,0)--(0,0);
\draw  (-1,1) node[boper] {};
\end{tikzpicture}
\;+x_r^{-1}\,
\begin{tikzpicture}[baseline=-3pt,scale=0.4]
\draw (2,1)  --  (-1,-2);
\draw  (1,2) -- (-2,-1);
\draw  (-1,2)--(2,-1);
\draw  (-2,1) --(1,-2);
\draw (0,0) node[blob] {};\draw (-2,0) node[blob] {};\draw (2,0) node[blob] {} ;
\draw (0,2) node[blob] {};\draw (0,-2) node[blob] {};
\draw (2,2) node[blob] {};\draw (2,-2) node[blob] {};
\draw (-2,2) node[blob] {};\draw (-2,-2) node[blob] {} ;
\draw[wavy=0.5](-4,0)--(0,0);
\draw  (1,1) node[boper] {};
\end{tikzpicture}
\label{eq:4term}\eec
\noindent Note that we have used the relation $X Z^{-1} = q^2 Z^{-1} X$ to rewrite the first term on the right-hand-side of this equation. 

In order to rewrite the relationship \eqref{eq:4term} in terms of CP weights, we need to understand the effect of the tail operator purely in terms of the  modification of CP weights. Recall that the action of the tail operator $t_0z_0^{-1}$ on the representation $V_{rr'}$ is given by
\be
\pi_{rr'}(t_0z_0^{-1}) = f_r f_{r'} Z^{-1} \,, \label{eq:tail}
\ee
where $Z^{-1}$ is the cyclic shift matrix that acts on canonical basis vectors $v_{a}$ ($a=0,1,\cdots,N-1$) as $Z^{-1} v_{a}=v_{a+1|_{\hbox{mod } N}}$.
Hence, we can identify
\bec
\begin{tikzpicture}[baseline=-3pt,scale=0.6]
\draw[bline] (-2,0) node[left] {$a$} -- (0,0) node[right] {$b$} ;
\draw[wavy=0.3] (-1,-1) -- (-1,1);
\draw (-2,0) node[bblob] {}; \draw (0,0) node[bblob] {};
\draw[dashed] (-1,-1) --  (-2,0) -- (-1,1) -- (0,0) -- cycle;
\draw (-1,-1) circle [radius=0.08];\draw (-1,1) circle [radius=0.08];
\end{tikzpicture}&=&\frac{f_r}{f_s}W_{rs}(a-b+1),\quad 
\begin{tikzpicture}[baseline=-3pt,scale=0.6]
\draw[bline] (-2,0) node[left] {$a$} -- (0,0) node[right] {$b$} ;
\draw[wavy=0.3] (-1,1) -- (-1,-1);
\draw (-2,0) node[bblob] {}; \draw (0,0) node[bblob] {};
\draw[dashed] (-1,-1) --  (-2,0) -- (-1,1) -- (0,0) -- cycle;
\draw (-1,-1) circle [radius=0.08];\draw (-1,1) circle [radius=0.08];
\end{tikzpicture}=\frac{f_s}{f_r}W_{rs}(a-b-1),\\[6mm]
\begin{tikzpicture}[baseline=-3pt,scale=0.6]
\draw[bline] (-1,1) node[above] {$a$} -- (-1,-1) node[below] {$b$} ;
\draw[wavy=0.3] (-2,0) -- (0,0);
\draw (-1,-1) node[bblob] {}; \draw (-1,1) node[bblob] {};
\draw[dashed] (-1,-1) --  (-2,0) -- (-1,1) -- (0,0) -- cycle;
\draw (-2,0) circle [radius=0.08];\draw (0,0) circle [radius=0.08];
\end{tikzpicture}&=&f_r f_s\bW_{rs}(a-b+1),\quad
\begin{tikzpicture}[baseline=-3pt,scale=0.6]
\draw[bline] (-1,1) node[above] {$a$} -- (-1,-1) node[below] {$b$} ;
\draw[wavy=0.3] (0,0) -- (-2,0);
\draw (-1,-1) node[bblob] {}; \draw (-1,1) node[bblob] {};
\draw[dashed] (-1,-1) --  (-2,0) -- (-1,1) -- (0,0) -- cycle;
\draw (-2,0) circle [radius=0.08];\draw (0,0) circle [radius=0.08];
\end{tikzpicture}=\frac{1}{f_r f_s}\bW_{rs}(a-b-1).
\label{eq:CPdisorder}\eec
In this way we identify the tail operator as a disorder operator for the 
CP model. 

The diagrammatic identity of \eqref{eq:4term} can be written as a relation
around any horizontal CP plaquette inserted into a larger partition function as
\begin{equation}
  \label{eq:dhhoriz1}
  \begin{aligned}
    & -y_r^{-1}\, \begin{tikzpicture}[baseline=15pt,scale=0.6]
      \draw (0,0) node[blob] {};\draw (0,2) node[blob] {};
      \draw (-1,1)  node[boper] {};
      \draw  (1,1) node[bblob] {};
      \draw[wavy=0.5](-3,0)--(-2,0) -- (-2,2) -- (0,2);
      \draw[bline] (-1,1) -- (1,1);
      \draw[dashed] (-1,1) --  (0,2) -- (1,1) -- (0,0) -- cycle;
    \end{tikzpicture}
    \;+ q^2 y_s^{-1}\hspace*{-4mm}
    \begin{tikzpicture}[baseline=15pt,scale=0.6]
      \draw (0,0) node[blob] {};\draw (0,2) node[blob] {};
      \draw (-1,1)  node[boper] {};
      \draw  (1,1) node[bblob] {};
      \draw[wavy=0.5](-3,0)--(0,0);
      \draw[bline] (-1,1) -- (1,1);
      \draw[dashed] (-1,1) --  (0,2) -- (1,1) -- (0,0) -- cycle;
    \end{tikzpicture} \\[6mm]
    & -x_r^{-1}\,
    \begin{tikzpicture}[baseline=15pt,scale=0.6]
      \draw (0,0) node[blob] {};\draw (0,2) node[blob] {};
      \draw (-1,1)  node[bblob] {};
      \draw  (1,1) node[boper] {};
      \draw[wavy=0.5](-3,0)--(0,0);
      \draw[bline] (-1,1) -- (1,1);
      \draw[dashed] (-1,1) --  (0,2) -- (1,1) -- (0,0) -- cycle;
    \end{tikzpicture}
    \;+x_s^{-1}\,
    \begin{tikzpicture}[baseline=15pt,scale=0.6]
      \draw (0,0) node[blob] {};\draw (0,2) node[blob] {};
      \draw (-1,1)  node[bblob] {};
      \draw  (1,1) node[boper] {};
      \draw[wavy=0.5](-3,0)--(-2,0) -- (-2,2) -- (0,2);
      \draw[bline] (-1,1) -- (1,1);
      \draw[dashed] (-1,1) --  (0,2) -- (1,1) -- (0,0) -- cycle;
    \end{tikzpicture}
    =0 \,.
  \end{aligned}
\end{equation}
Consider the following non-local operator $j_{\bar{e}_0}(\vec r)$ defined in terms of the above half-currents by
\begin{equation*}
  j_{\bar{e}_0}\left( \frac{\vec r_\sigma+\vec r_\mu}{2} \right)=T[\mu_{\bar e_0}(\vec r_\mu)\sigma(\vec r_\sigma)] \,,
\end{equation*}
where
\begin{itemize}
\item[-] $\sigma(\vec r_\sigma)$ corresponds to the insertion of $X=\blacksquare$ at embedded CP site $\vec r_\sigma$
\item[-]$\mu_{\bar e_0}(\vec r_\mu)$ is the above tail/disorder operator 
  ending at embedded dual CP site $\vec r_\mu$
\item[-] $T$ is `tail ordering' defined as:
\begin{align*}
  & T[\mu_{\bar e_0}(\vec r_\mu) \sigma_1(\vec r_\sigma)] = \\
  & \qquad  \hbox{quasi-local op. $\mu_{\bar e_0}(\vec r_\mu) \sigma(\vec r_\sigma)$ with tail}
  \begin{cases}
    & \hbox{locally above $\vec r_\sigma$  if  $\hbox{Im}(z_\mu) \leq \hbox{Im}(z_\sigma)$} \\
    & \hbox{locally below $\vec r_\sigma$  if  $\hbox{Im}(z_\mu)   >  \hbox{Im}(z_\sigma)$}
  \end{cases}
\end{align*}
\end{itemize}

With this definition, the graphical relation~\eqref{eq:dhhoriz1} can be written simply as
\begin{equation} \label{eq:dhhoriz2}
  -y_r^{-1} j_{\bar{e}_0}(\vec r_1) + q^2 y_s^{-1} j_{\bar{e}_0}(\vec r_2)
  - x_r^{-1} j_{\bar{e}_0}(\vec r_3) + x_s^{-1} j_{\bar{e}_0}(\vec r_4) = 0 \,,
\end{equation}
where we have denoted the mid-edges of the plaquette as follows:
\begin{equation}
  \label{eq:hplaquette}
  \begin{tikzpicture}[baseline=20pt,scale=0.8]
    \draw (0,0) node[blob] {};\draw (0,2) node[blob] {};
    \draw (-1,1)  node[bblob] {};
    \draw  (1,1) node[bblob] {};
    \draw (-0.5,1.5) node {$*$};
    \draw (-0.5,0.5) node {$*$};
    \draw (0.5,0.5) node {$*$};
    \draw (0.5,1.5) node {$*$};
    \draw (-0.5,1.6) node[left] {$\vec r_1$};
    \draw (-0.5,0.4) node[left] {$\vec r_2$};
    \draw (0.5,0.4) node[right] {$\vec r_3$};
    \draw (0.5,1.6) node[right] {$\vec r_4$};
    \draw[bline] (-1,1) -- (1,1);
    \draw[dashed] (-1,1) --  (0,2) -- (1,1) -- (0,0) -- cycle;
  \end{tikzpicture} \,.
\end{equation}

\subsection{The operator $\cO_{\bar{e}_0}^{(s)}$ and twisted Cauchy-Riemann equation}
\label{sec:def-O}


Let us parameterise $(x,y,\mu)$ in terms of  $(u,\phi,\bar{\phi})$ as in~\eqref{eq:param}, and introduce
\begin{align}
  &\alpha_1 = \frac{u_s-u_r}{2}-\pi \,,
  \quad &&\alpha_2 = -\frac{u_s-u_r}{2}+\pi \,,
  \quad &&\alpha_3 = \frac{u_s-u_r}{2} \,,
  \quad &&\alpha_4 = -\frac{u_s-u_r}{2} \,.
\end{align}
The linear relation~\eqref{eq:dhhoriz2} reads:
\begin{equation}
  e^{i(\phi_r+\alpha_1)/N} j_{\bar{e}_0}(\vec r_1) - e^{i(\phi_s+\alpha_2)/N} j_{\bar{e}_0}(\vec r_2)
  + e^{i(-\phi_r+\alpha_3)/N} j_{\bar{e}_0}(\vec r_3) - e^{i(-\phi_s+\alpha_4)/N} j_{\bar{e}_0}(\vec r_4) = 0 \,.
\end{equation}
We choose $\theta=u_s-u_r$ as the embedding angle in~\eqref{eq:embed-graph}, so that the quantity $\alpha_j$ coincides with the principal argument of $(z_\sigma-z_\mu)$ on the corresponding edge of the plaquette. Then the linear relation takes the form of a ``twisted Cauchy-Riemann'' relation:
\begin{equation} \label{eq:contour}
  e^{\frac{i\phi_r}{N}} \delta z_1\ \cO^{(s)}_{\bar{e}_0}(\vec r_1)
  + e^{\frac{i\phi_s}{N}} \delta z_2\ \cO^{(s)}_{\bar{e}_0}(\vec r_2)
  + e^{-\frac{i\phi_r}{N}} \delta z_3\ \cO^{(s)}_{\bar{e}_0}(\vec r_3)
  + e^{-\frac{i\phi_s}{N}} \delta z_4\ \cO^{(s)}_{\bar{e}_0}(\vec r_4) = 0 \,,
\end{equation}
where we have introduced the lattice current
\begin{equation}
  \cO^{(s)}_{\bar{e}_0}(\vec r) = \exp[-is \alpha(\vec r)] \ j_{\bar{e}_0}(\vec r) \,,
\end{equation}
and where $\alpha(\vec r)$ is the principal argument of the oriented edge $(z_\sigma-z_\mu)$ carrying the point $\vec r$, and the spin $s$ is set to $s=1-1/N$.

\subsection{The $\bW_{rs}$ plaquette}
Equation \eqref{eq:contour} is a discrete integral relation around a  $W_{rs}$ plaquette. There is a direct extension of the above arguments that leads to a discrete integral relation for $\cO_{\bar{e}_0}(z)$ around a $\bW_{rs}$ plaquette. The starting point is to consider the commutation relation \eqref{eq:suffcond3}, that is
\be ( T_{rs}\ot 1)   [\pi_{sr}\ot \pi_{sr}(\Delta(\bar{e}_0))]&=&  
[\pi_{rs}\ot \pi_{sr}(\Delta(\bar{e}_0))]( T_{rs}\ot 1)\label{eq:suffcond31} \,.
\ee
Using the splitting into half-currents given by \eqref{eq:be0graph}, this can be represented by
\ben
&&\hspace*{-15mm}x_r^{-1}\,
\begin{tikzpicture}[baseline=-3pt,scale=0.4]
\draw[aline] (2,1) node[right] {$s$} --  (-1,-2);
\draw[aline]  (1,2) node[above] {$s$} -- (-2,-1);
\draw[aline]  (-1,2) node[above] {$r$}--(2,-1);
\draw[aline]  (-2,1) node[left] {$r$} --(1,-2);
\draw (0,0) node[blob] {};\draw (-2,0) node[blob] {};\draw (2,0) node[blob] {} ;
\draw (0,2) node[blob] {};\draw (0,-2) node[blob] {};
\draw (2,2) node[blob] {};\draw (2,-2) node[blob] {};
\draw (-2,2) node[blob] {};\draw (-2,-2) node[blob] {} ;
\draw[wavy=0.5](-4,0) -- (-2,0);
\draw  (-1,1) node[boper] {};
\end{tikzpicture}
-y_s^{-1}\,
\begin{tikzpicture}[baseline=-3pt,scale=0.4]
\draw (2,1)  --  (-1,-2);
\draw  (1,2) -- (-2,-1);
\draw  (-1,2)--(2,-1);
\draw  (-2,1) --(1,-2);
\draw (0,0) node[blob] {};\draw (-2,0) node[blob] {};\draw (2,0) node[blob] {} ;
\draw (0,2) node[blob] {};\draw (0,-2) node[blob] {};
\draw (2,2) node[blob] {};\draw (2,-2) node[blob] {};
\draw (-2,2) node[blob] {};\draw (-2,-2) node[blob] {} ;
\draw[wavy=0.5](-4,0) -- (-2,0) -- (-2,2) -- (0,2) -- (0,0);
\draw  (-1,1) node[boper] {};
\end{tikzpicture}
+x_r^{-1}\,
\begin{tikzpicture}[baseline=-3pt,scale=0.4]
\draw (2,1)  --  (-1,-2);
\draw  (1,2) -- (-2,-1);
\draw  (-1,2)--(2,-1);
\draw  (-2,1) --(1,-2);
\draw (0,0) node[blob] {};\draw (-2,0) node[blob] {};\draw (2,0) node[blob] {} ;
\draw (0,2) node[blob] {};\draw (0,-2) node[blob] {};
\draw (2,2) node[blob] {};\draw (2,-2) node[blob] {};
\draw (-2,2) node[blob] {};\draw (-2,-2) node[blob] {} ;
\draw[wavy=0.5](-4,0) -- (0,0);
\draw  (1,1) node[boper] {};
\end{tikzpicture}
-y_s^{-1}\,
\begin{tikzpicture}[baseline=-3pt,scale=0.4]
\draw (2,1)  --  (-1,-2);
\draw  (1,2) -- (-2,-1);
\draw  (-1,2)--(2,-1);
\draw  (-2,1) --(1,-2);
\draw (0,0) node[blob] {};\draw (-2,0) node[blob] {};\draw (2,0) node[blob] {} ;
\draw (0,2) node[blob] {};\draw (0,-2) node[blob] {};
\draw (2,2) node[blob] {};\draw (2,-2) node[blob] {};
\draw (-2,2) node[blob] {};\draw (-2,-2) node[blob] {} ;
\draw[wavy=0.5](-4,0) -- (0,0) -- (0,2) -- (2,2) -- (2,0);
\draw  (1,1) node[boper] {};
\end{tikzpicture}
\\[3mm]
&&\hspace*{-15mm}=
x_s^{-1}\,
\begin{tikzpicture}[baseline=-3pt,scale=0.4]
\draw (2,1)  --  (-1,-2);
\draw  (1,2)   -- (-2,-1);
\draw  (-1,2) --(2,-1);
\draw  (-2,1)  --(1,-2);
\draw (0,0) node[blob] {};\draw (-2,0) node[blob] {};\draw (2,0) node[blob] {} ;
\draw (0,2) node[blob] {};\draw (0,-2) node[blob] {};
\draw (2,2) node[blob] {};\draw (2,-2) node[blob] {};
\draw (-2,2) node[blob] {};\draw (-2,-2) node[blob] {} ;
\draw[wavy=0.5](-4,0)--(-2,0);
\draw  (-1,-1) node[boper] {};
\end{tikzpicture}
-y_r^{-1}\,
\begin{tikzpicture}[baseline=-3pt,scale=0.4]
\draw (2,1)  --  (-1,-2);
\draw  (1,2) -- (-2,-1);
\draw  (-1,2)--(2,-1);
\draw  (-2,1) --(1,-2);
\draw (0,0) node[blob] {};\draw (-2,0) node[blob] {};\draw (2,0) node[blob] {} ;
\draw (0,2) node[blob] {};\draw (0,-2) node[blob] {};
\draw (2,2) node[blob] {};\draw (2,-2) node[blob] {};
\draw (-2,2) node[blob] {};\draw (-2,-2) node[blob] {} ;
\draw[wavy=0.5](-4,0) -- (0,0);
\draw  (-1,-1) node[boper] {};
\end{tikzpicture}
+x_r^{-1}\,
\begin{tikzpicture}[baseline=-3pt,scale=0.4]
\draw (2,1)  --  (-1,-2);
\draw  (1,2) -- (-2,-1);
\draw  (-1,2)--(2,-1);
\draw  (-2,1) --(1,-2);
\draw (0,0) node[blob] {};\draw (-2,0) node[blob] {};\draw (2,0) node[blob] {} ;
\draw (0,2) node[blob] {};\draw (0,-2) node[blob] {};
\draw (2,2) node[blob] {};\draw (2,-2) node[blob] {};
\draw (-2,2) node[blob] {};\draw (-2,-2) node[blob] {} ;
\draw[wavy=0.5](-4,0)--(0,0);
\draw  (1,1) node[boper] {};
\end{tikzpicture}
-y_s^{-1}\,
\begin{tikzpicture}[baseline=-3pt,scale=0.4]
\draw (2,1)  --  (-1,-2);
\draw  (1,2) -- (-2,-1);
\draw  (-1,2)--(2,-1);
\draw  (-2,1) --(1,-2);
\draw (0,0) node[blob] {};\draw (-2,0) node[blob] {};\draw (2,0) node[blob] {} ;
\draw (0,2) node[blob] {};\draw (0,-2) node[blob] {};
\draw (2,2) node[blob] {};\draw (2,-2) node[blob] {};
\draw (-2,2) node[blob] {};\draw (-2,-2) node[blob] {} ;
\draw[wavy=0.5](-4,0) -- (0,0) -- (0,2) -- (2,2) -- (2,0);
\draw  (1,1) node[boper] {};
\end{tikzpicture}
\een 
Cancelling the common terms, and writing in terms of the $\ol W$ plaquette gives
\be
&&x_r^{-1} \begin{tikzpicture}[baseline=-3pt,scale=0.6]
\draw[bline] (-1,1)  -- (-1,-1);
\draw[wavy=0.3] (-4,0) -- (-2,0);
\draw (-1,-1) node[bblob] {}; \draw (-1,1) node[oper] {};
\draw[dashed] (-1,-1) --  (-2,0) -- (-1,1) -- (0,0) -- cycle;
\draw (-2,0) circle [radius=0.08];\draw (0,0) circle [radius=0.08];
\end{tikzpicture}\quad -q^2 y_s^{-1}\quad
\begin{tikzpicture}[baseline=-3pt,scale=0.6]
\draw[bline] (-1,1)  -- (-1,-1);
\draw[wavy=0.3] (-4,0) -- (0,0);
\draw (-1,-1) node[bblob] {}; \draw (-1,1) node[oper] {};
\draw[dashed] (-1,-1) --  (-2,0) -- (-1,1) -- (0,0) -- cycle;
\draw (-2,0) circle [radius=0.08];\draw (0,0) circle [radius=0.08];
\end{tikzpicture}\\
&&-x_s^{-1}
 \begin{tikzpicture}[baseline=-3pt,scale=0.6]
\draw[bline] (-1,1)  -- (-1,-1);
\draw[wavy=0.3] (-4,0) -- (-2,0);
\draw (-1,-1) node[oper] {}; \draw (-1,1) node[bblob] {};
\draw[dashed] (-1,-1) --  (-2,0) -- (-1,1) -- (0,0) -- cycle;
\draw (-2,0) circle [radius=0.08];\draw (0,0) circle [radius=0.08];
\end{tikzpicture}\quad + y_r^{-1}\quad
\begin{tikzpicture}[baseline=-3pt,scale=0.6]
\draw[bline] (-1,1)  -- (-1,-1);
\draw[wavy=0.3] (-4,0) -- (0,0);
\draw (-1,-1) node[oper] {}; \draw (-1,1) node[bblob] {};
\draw[dashed] (-1,-1) --  (-2,0) -- (-1,1) -- (0,0) -- cycle;
\draw (-2,0) circle [radius=0.08];\draw (0,0) circle [radius=0.08];
\end{tikzpicture}
\quad =0 \,.\ee
Defining the operator $\cO^{(s)}_{\bar{e}_0}(\vec r)$ exactly as above, this leads to the discrete
integral condition 
\begin{equation} \label{eq:contour2}
  e^{-\frac{i\phi_r}{N}} \delta z_1 \cO^{(s)}_{\bar{e}_0}(\vec r_1)
  + e^{-\frac{i\phi_s}{N}} \delta z_2 \cO^{(s)}_{\bar{e}_0}(\vec r_2)
  + e^{\frac{i\phi_r}{N}} \delta z_3 \cO^{(s)}_{\bar{e}_0}(\vec r_3)
  + e^{\frac{i\phi_s}{N}} \delta z_4 \cO^{(s)}_{\bar{e}_0}(\vec r_4) = 0 \,,
\end{equation}
around the embedded plaquette
\bec
\label{eq:vplaquette}
\begin{tikzpicture}[baseline=20pt,scale=0.8]
  \draw (0,0) node[bblob] {};\draw (0,2) node[bblob] {};
  \draw (-1,1)  node[blob] {};
  \draw  (1,1) node[blob] {};
  \draw (-0.5,1.5) node {$*$};
  \draw (-0.5,0.5) node {$*$};
  \draw (0.5,0.5) node {$*$};
  \draw (0.5,1.5) node {$*$};
  \draw (-0.5,1.6) node[left] {$\vec r_1$};
  \draw (-0.5,0.4) node[left] {$\vec r_2$};
  \draw (0.5,0.4) node[right] {$\vec r_3$};
  \draw (0.5,1.6) node[right] {$\vec r_4$};
  \draw[bline] (0,2) -- (0,0);
  \draw[dashed] (-1,1) --  (0,2) -- (1,1) -- (0,0) -- cycle;
\end{tikzpicture} \,.
\eec
The coefficients appearing in \eqref{eq:contour} and \eqref{eq:contour2} are such that when we consider any larger region, the contribution from internal points cancel and we are left with a discretely holomorphicity relation expressed solely on the boundary of the region.  

\subsection{Other currents}
It is possible to define quasi-local operators in terms of
the half-currents associated with the other generators. We consider these in turn.

\subsubsection{The current for $e_0$}
The coproduct and representation of $e_0$ are given by 
\be
\Delta(e_0)&=& e_0 \otimes \id + t_0 z_0 \otimes e_0,\quad
\Delta(t_0 z_0)= t_0 z_0\ot t_0 z_0 \,, \label{eq:coprod3}\\
\pi_{rr'}(e_0)&=& 
\beta X^{-1}\left[x_{r'}-y_r\pi_{rr'}(t_0z_0) \right],\quad
\pi_{rr'}(t_0z_0) = \frac{1}{\mu_r \mu_{r'}} Z^{-1}\, ,
\nn
\ee
where $\beta:=-q/(q^2-1)^2$.
Proceeding as above, may now define a new half-current
\ben
j_{e_0}((\vec r_\sigma+\vec r_\mu)/2)=T(\mu_{e_0}(\vec r_\mu) \sigma^\dag(\vec r_\sigma)),
\een
where
\begin{itemize}
\item[-] $\sigma^\dag(\vec r_\sigma)$ corresponds to the insertion of $X^{-1}$ at embedded CP site $\vec r_\sigma$
\item[-] $\mu_{e_0}(\vec r_\mu)$ is a new disorder operator corresponding to the insertion of the operator $t_0z_0$ along a path ending at the dual CP site $\vec r_\mu$. The effect of this disorder operator is similar to that of \eqref{eq:CPdisorder} except that $f_{r,s}\rightarrow \mu_{r,s}^{-1}$ on the right-hand side. 
\end{itemize} 

Defining $\cO_{e_o}^{(-s)}(\vec r)= \exp[+is\alpha(\vec r)] j_{e_0}(\vec r)$, we can follow the previous analysis to arrive at the discrete antiholomorphicity conditions
\begin{align*}
  e^{-\frac{i\phi_r}{N}} \delta \bar{z}_1 \cO^{(-s)}_{e_o}(\vec r_1)
  + e^{-\frac{i\phi_s}{N}} \delta \bar{z}_2 \cO^{(-s)}_{e_o}(\vec r_2)
  + e^{\frac{i\phi_r}{N}} \delta \bar{z}_3 \cO^{(-s)}_{e_o}(\vec r_3)
  + e^{\frac{i\phi_s}{N}} \delta \bar{z}_4 \cO^{(-s)}_{e_o}(\vec r_4) &= 0 \,, \\[3mm]
  e^{\frac{i\phi_r}{N}} \delta \bar{z}_1 \cO^{(-s)}_{e_o}(\vec r_1)
  + e^{\frac{i\phi_s}{N}} \delta \bar{z}_2 \cO^{(-s)}_{e_o}(\vec r_2)
  + e^{-\frac{i\phi_r}{N}} \delta \bar{z}_3 \cO^{(-s)}_{e_o}(\vec r_3)
  + e^{-\frac{i\phi_s}{N}} \delta \bar{z}_4 \cO^{(-s)}_{e_o}(\vec r_4) &= 0 \,,
\end{align*}
around the $W$ plaquette \eqref{eq:hplaquette} and $\overline{W}$ plaquette \eqref{eq:vplaquette} respectively.

\subsubsection{The current for $\bar{e}_1$}
A similar consideration leads us to define
\ben 
j_{\bar{e}_1}((\vec r_\sigma+\vec r_\mu)/2)=T(\mu_{\bar{e}_1}(\vec r_\mu) \sigma^\dag(\vec r_\sigma)) \,,
\een
where
\begin{itemize}
\item[-] $\sigma^\dag(\vec r_\sigma)$ corresponds to the insertion of $X^{-1}$ at embedded CP site $\vec r_\sigma$
\item[-]$\mu_{\bar{e}_1}(\vec r_\mu)$ is a new disorder operator corresponding to the insertion of the operator $t_1z_1^{-1}$ along a path ending at the dual CP site $\vec r_\mu$. Noting that
  \ben
  \pi_{rr'}(t_1 z_1^{-1})=\frac{1}{f_r f_r'} Z
  \een
and comparing to equation \eqref{eq:tail}, we see that this $\mu_{\bar{e}_1}(\vec r_\mu)$ disorder operator has the same action as in \eqref{eq:CPdisorder}, but now with the arrow is directed outwards from the dual CP site $\vec r_\mu$.

\end{itemize} 
Defining  $\cO_{\bar{e}_1}^{(s)}(\vec r)= \exp[-is\alpha(\vec r)] j_{\bar{e}_1}(\vec r)$, we obtain the $W$ and $\overline{W}$ discrete holomorphicity conditions
\begin{align}
  e^{-\frac{i\phi_r}{N}} \delta z_1 \cO^{(s)}_{\bar{e}_1}(\vec r_1)
  + e^{-\frac{i\phi_s}{N}} \delta z_2 \cO^{(s)}_{\bar{e}_1}(\vec r_2)
  + e^{\frac{i\phi_r}{N}} \delta z_3 \cO^{(s)}_{\bar{e}_1}(\vec r_3)
  + e^{\frac{i\phi_s}{N}} \delta z_4 \cO^{(s)}_{\bar{e}_1}(\vec r_4) &= 0 \,, \label{eq:e1barhoriz} \\[3mm]
  e^{\frac{i\phi_r}{N}} \delta z_1 \cO^{(s)}_{\bar{e}_1}(\vec r_1)
  + e^{\frac{i\phi_s}{N}} \delta z_2 \cO^{(s)}_{\bar{e}_1}(\vec r_2)
  + e^{-\frac{i\phi_r}{N}} \delta z_3 \cO^{(s)}_{\bar{e}_1}(\vec r_3)
  + e^{-\frac{i\phi_s}{N}} \delta z_4 \cO^{(s)}_{\bar{e}_1}(\vec r_4) &= 0 \,.
\end{align}

\subsubsection{The current for $e_1$}
Finally, by considering the half-currents that make up $e_1$, we arrive at the following 
definition of a quasi-local operator
$$
j_{e_1}((\vec r_\sigma+\vec r_\mu)/2)=T(\mu_{e_1}(\vec r_\mu) \sigma(\vec r_\sigma)) \,,
$$
where
\begin{itemize}
\item[-] $\sigma(\vec r_\sigma)$ corresponds to the insertion of $X$ at embedded CP site $\vec r_\sigma$
\item[-]$\mu_{e_1}(\vec r_\mu)$ is a disorder operator corresponding to the insertion of the operator $t_1z_1$
along a path ending at the dual CP site $\vec r_\mu$. This effect of the disorder operator is given by \eqref{eq:CPdisorder} with $f_{r,s}\rightarrow \mu_{r,s}^{-1}$ on the right-hand side and with the arrow leaving $\vec r_\mu$.
\end{itemize} 
Defining  $\cO^{(-s)}_{e_1}(\vec r)= \exp[is\alpha(\vec r)] j_{e_1}(\vec r)$, we obtain the $W$ and $\overline{W}$ discrete anti-holomorphicity conditions
\begin{align*}
  e^{\frac{i\phi_r}{N}} \delta \bar{z}_1 \cO^{(-s)}_{e_1}(\vec r_1)
  + e^{\frac{i\phi_s}{N}} \delta \bar{z}_2 \cO^{(-s)}_{e_1}(\vec r_2)
  + e^{-\frac{i\phi_r}{N}} \delta \bar{z}_3 \cO^{(-s)}_{e_1}(\vec r_3)
  + e^{-\frac{i\phi_s}{N}} \delta \bar{z}_4 \cO^{(-s)}_{e_1}(\vec r_4) &= 0 \,, \\[3mm]
  e^{-\frac{i\phi_r}{N}} \delta \bar{z}_1 \cO^{(-s)}_{e_1}(\vec r_1)
  + e^{-\frac{i\phi_s}{N}} \delta \bar{z}_2 \cO^{(-s)}_{e_1}(\vec r_2)
  + e^{\frac{i\phi_r}{N}} \delta \bar{z}_3 \cO^{(-s)}_{e_1}(\vec r_3)
  + e^{\frac{i\phi_s}{N}} \delta \bar{z}_4 \cO^{(-s)}_{e_1}(\vec r_4) &= 0 \,.
\end{align*}

In summary, $\bar{e}_i$ yield half-currents with discrete holomorphicity and spin $s=(1-1/N)$, whilst $e_i$ yield ones with discrete antiholomorphicity and spin $-s$. The coefficients in the discrete relations around either $(\vec r_1,\vec r_2,\vec r_3,\vec r_4)$ plaquette are just $(e^{\pm i\phi_r/N}, e^{\pm i\phi_s/N},e^{\mp i\phi_r/N},e^{\mp i\phi_r/N})$ in all cases. 

\section{Physical interpretation}
\label{sec:phys}

In this section, we discuss the physical meaning of the linear relations derived above for the operators $\cO^{(s)}_{\bar e_i}$ and $\cO^{(-s)}_{e_i}$. For simplicity, we restrict the discussion to the case of $\cO^{(s)}_{\bar e_0}$ and the linear relation~\eqref{eq:contour} around a $W_{rs}$ plaquette, but very similar results hold for the other cases. Also, in order to lighten notation, we drop the indices $\bar e_0$ in $j_{\bar e_0}(\vec r)$ or $\cO^{(s)}_{\bar e_0}(\vec r)$, and simply write $j(\vec r)$ and $\cO^{(s)}(\vec r)$.

\subsection{Discrete linear relation in the vicinity of the FZ point}

At the  FZ point $\phi_{r,s}=\bar{\phi}_{r,s}=0$, \eqref{eq:contour} takes the simple form
\begin{equation} \label{eq:CR}
  \delta z_1 \cO^{(s)}(\vec r_1) + \delta z_2 \cO^{(s)}(\vec r_2)
  + \delta z_3 \cO^{(s)}(\vec r_3) + \delta z_4 \cO^{(s)}(\vec r_4) = 0 \,.
\end{equation}
Note that at this point $\cO^{(s)}(z)$ coincides with the lattice parafermion of~\cite{RajCardy},
and \eqref{eq:CR} is the discrete Cauchy-Riemann relation of the form $\bar\partial \psi_s = 0$ found empirically in~\cite{RajCardy}.

In the vicinity of the FZ point, using \eqref{eq:Ck2}, we can write the linearised relations
\begin{equation}
  \phi_r = \cos\theta\ \phi_s + i\sin\theta\ \bar\phi_s + O(\phi_s^3,\bar\phi_s^3) \,,
  \qquad
  \bar\phi_r = i\sin\theta\ \phi_s + \cos\theta\ \bar\phi_s + O(\phi_s^3,\bar\phi_s^3) \,.
\end{equation}
Hence, if we introduce the notation $\phi^{\pm} = (\phi \pm \bar\phi)/2$, we get $\phi_r^\pm \sim e^{\pm i\theta} \phi_s^\pm$, and \eqref{eq:contour} becomes
\begin{align}
  & \delta z_1 \cO^{(s)}(\vec r_1) + \delta z_2 \cO^{(s)}(\vec r_2)
  + \delta z_3 \cO^{(s)}(\vec r_3) + \delta z_4 \cO^{(s)}(\vec r_4) = \nn \\
  & \qquad - \alpha_s^+ \left[
    t \cO^{(s)}(\vec r_1) + t^{-1} \cO^{(s)}(\vec r_2)
    + t \cO^{(s)}(\vec r_3) + t^{-1} \cO^{(s)}(\vec r_4)
  \right] \nn \\
  & \qquad + \alpha_s^- \left[
    t^{-1} \cO^{(s-2)}(\vec r_1) + t \cO^{(s-2)}(\vec r_2)
    + t^{-1} \cO^{(s-2)}(\vec r_3) + t \cO^{(s-2)}(\vec r_4)
  \right] \nn \\
  & \qquad +i (\alpha_s^+)^2 \left[
    t \cO^{(s-1)}(\vec r_1) + t^{-1} \cO^{(s-1)}(\vec r_2)
    + t \cO^{(s-1)}(\vec r_3) + t^{-1} \cO^{(s-1)}(\vec r_4)
  \right]  \nn \\
  & \qquad -i (\alpha_s^-)^2 \left[
    t^{-1} \cO^{(s-1)}(\vec r_1) + t \cO^{(s-1)}(\vec r_2)
    + t^{-1} \cO^{(s-1)}(\vec r_3) + t \cO^{(s-1)}(\vec r_4)
  \right] \nn \\ &\qquad +O((\alpha_s^{\pm})^3), \label{eq:CR2}
\end{align}
where we have set
\begin{equation}
  t= -ie^{i\theta} \,,
  \qquad \alpha_s^+ = \frac{e^{i\theta/2} \phi_s^+}{N} \,,
  \qquad \alpha_s^- = \frac{e^{-i\theta/2} \phi_s^-}{N} \,.
\end{equation}
Note that the sum of coefficients multiplying the $\cO$'s in each bracket on the RHS of~\eqref{eq:CR2} is $4\sin\theta$, which is proportional to the area of the plaquette. In the case $\theta=\pi/2$, then $t=1$ and \eqref{eq:CR2} takes the simpler form:
\begin{align}
  & \delta z_1 \cO^{(s)}(\vec r_1) + \delta z_2 \cO^{(s)}(\vec r_2)
  + \delta z_3 \cO^{(s)}(\vec r_3) + \delta z_4 \cO^{(s)}(\vec r_4) = \nn \\
  & \qquad - \alpha_s^+ \widehat{\cO}^{(s)}(\vec p)
  + \alpha_s^- \widehat{\cO}^{(s-2)}(\vec p)
  +i [(\alpha_s^+)^2-(\alpha_s^-)^2] \widehat{\cO}^{(s-1)}(\vec p)
  +O((\alpha_s^{\pm})^3) \,, \label{eq:CR3}
\end{align}
where we have defined $\vec p$ as the center of the plaquette, and
$$
\widehat{\cO}(\vec p) = \cO(\vec r_1) + \cO(\vec r_2) + \cO(\vec r_3) + \cO(\vec r_4) \,.
$$

\subsection{Comparison with perturbed CFT}

In general, consider the perturbed action~\cite{Zam89}
$$
S = S_{\rm CFT} + g \int d^2r \ \Phi_{h,\bar h}(z,\bar z) \,,
$$
where $S_{\rm CFT}$ is the action of some CFT, and $\Phi_{h,\bar h}$ is an operator in the spectrum of this CFT. Suppose the unperturbed theory possesses a holomorphic current $\psi_s(z)$ with conformal spin $s$, and moreover, assume an OPE between the current and the perturbing field of the form:
\begin{equation} \label{eq:OPE}
  \psi_s(z) \Phi_{h,\bar h}(w,\bar w) = \dots + \frac{\chi(w,\bar w)}{z-w} + \dots
\end{equation}
By dimensional analysis, $\chi$ must have conformal weights $h_\chi = s+h-1$ and $\bar h_\chi = \bar h$. 
In the perturbed theory, $\psi_s$ is no longer a conserved current. More precisely, using the identity $\partial_{\bar z_1}[{1}/{(z_1-z_2)}] = \pi \delta(\vec r_1-\vec r_2)$, one gets at first order in $g$:
\begin{equation} \label{eq:dbar}
  \bar\partial \psi_s(z,\bar z) = \pi g \ \chi(z,\bar z) \,.
\end{equation}
So this simple line of arguments tells us how the conservation equation for the current $\psi_s$ is modified by the perturbation.
\bigskip

Let us now specialise to the $\Z_N$ parafermionic CFT for $N \geq 3$ (the case $N=2$ is treated separately in the next section). The chiral Potts model corresponds~\cite{Cardy93,Watts98} to a perturbation of this CFT by the energy operator $\ep$, together with the leading spin $\pm 1$ operators:
\begin{equation}
  S = S_{\rm FZ} + \int d^2r \left[
    \delta_+ \Phi_+(z,\bar z) + \delta_- \Phi_-(z,\bar z) + \tau \ep(z,\bar z)
  \right] \,.
\end{equation}
The energy operator $\ep$ has dimensions $h_\ep=\bar h_\ep=2/(N+2)$, and the spin $\pm 1$ operators (which are actually the descendants $W_{-1}\ep$ and $\ol W_{-1}\ep$ in terms of the underlying $W_N$ algebra) have dimensions $(h_{\Phi_+}, \bar h_{\Phi_+})=(h_\ep+1, h_\ep)$ and $(h_{\Phi_-},\bar h_{\Phi_-})=(h_\ep, h_\ep+1)$.
The parafermion of $\Z_N$ charges $m=\ol m=1$ has conformal spin $s=1-1/N$, and its conservation equation is modified by the perturbation according to~\eqref{eq:dbar}:
\begin{equation} \label{eq:dbar2}
  \bar\partial \psi_s(z,\bar z) = 
  \pi \delta_+ \ \chi_+(z,\bar z) + \pi \delta_- \ \chi_-(z,\bar z) + \pi \tau \ \chi_0(z,\bar z) \,,
 \end{equation}
where the operators on the right-hand side have $\Z_N$ charges $m=\ol m=1$, and conformal dimensions:
\begin{equation*}
  (h_{\chi_+}, \bar h_{\chi_+}) = (h_\ep+s, \bar h_\ep) \,,
  \quad
  (h_{\chi_-}, \bar h_{\chi_-}) = (h_\ep+s-1, \bar h_\ep+1) \,,
  \quad
  (h_{\chi_0}, \bar h_{\chi_0}) = (h_\ep+s-1, h_\ep) \,.
\end{equation*}
Their conformal spins are thus $s$, $(s-2)$ and $(s-1)$, respectively.

Hence, we see that we can interpret \eqref{eq:CR3} as a discrete version of the perturbed current conservation equation \eqref{eq:dbar2}, with the parameters in the latter related by
\begin{equation}
 \tau \propto (\delta_+^2 - \delta_-^2) \,.
\end{equation}

\subsection{The Ising case}

To make contact with previous work~\cite{RivaCardy}, it is convenient to introduce the ``bare'' current
\begin{equation}
  J = Z \otimes Z \otimes \dots \otimes Z \otimes X \,.
\end{equation}
For this operator, the linear relations \eqref{eq:contour} and \eqref{eq:e1barhoriz} arising from $j_{\bar e_0}$ and $j_{\bar e_1}$, when written in terms of the Ising parameterisation (\ref{eq:embed2}--\ref{eq:param-Ising2}), read respectively:
\begin{align}
  \Theta_1(\beta_r) H(\beta_s)J(\vec r_1) + \Theta_1(\beta_s) H(\beta_r)J(\vec r_2)
  + \Theta(\beta_r) H_1(\beta_s)J(\vec r_3) - \Theta(\beta_s) H_1(\beta_r)J(\vec r_4) &=0 \,, \label{eq:CRbare1} \\
  \Theta(\beta_r) H_1(\beta_s)J(\vec r_1)+ \Theta(\beta_s) H_1(\beta_r)J(\vec r_2)
  - \Theta_1(\beta_r) H(\beta_s)J(\vec r_3) + \Theta_1(\beta_s) H(\beta_r)J(\vec r_4) &=0  \,. \label{eq:CRbare2} 
\end{align}
When performing the $p \to 0$ expansion using~\eqref{eq:elliptic}, the terms of order $p^{1/2}$ vanish in the combination $\eqref{eq:CRbare1} -i \eqref{eq:CRbare2}$, and we obtain:
\begin{align}
  &\delta z_1\ \psi(\vec r_1) + \delta z_2\ \psi(\vec r_2) + \delta z_3\ \psi(\vec r_3) + \delta z_4\ \psi(\vec r_4)
  = \nn \\
  &\qquad\qquad -ip \left[ t^{-1} \bar\psi(\vec r_1) + t \bar\psi(\vec r_2) + t^{-1} \bar\psi(\vec r_3) + t \bar\psi(\vec r_4)
  \right] \,, \label{eq:CR-Ising}
\end{align}
where $t=-ie^{i\theta}$, and we have defined
\begin{equation}
  \psi(\vec r) = e^{-i\alpha(\vec r)/2} J(\vec r) \,,
  \qquad
  \bar\psi(\vec r) = e^{+i\alpha(\vec r)/2} J(\vec r) \,.
\end{equation}
In the RHS of~\eqref{eq:CR-Ising}, the sum of coefficients multiplying the $\bar\psi$'s is $(-4ip \sin\theta)$. Since the area of the plaquette is $\sin\theta$, the relation~\eqref{eq:CR-Ising} is thus a discrete version of the massive Dirac equation with mass $m=4p \sim k^2/4$:
\begin{equation}
  \bar\partial \psi = -im \ \bar\psi \,.
\end{equation}
Moreover, note that at $\theta=\pi/2$ we have $t=1$, we recover the simple form of~\cite{RivaCardy}:
\begin{equation}
  \delta z_1\ \psi(\vec r_1) + \delta z_2\ \psi(\vec r_2) + \delta z_3\ \psi(\vec r_3) + \delta z_4\ \psi(\vec r_4)
  = -ip \left[ \bar\psi(\vec r_1) + \bar\psi(\vec r_2) + \bar\psi(\vec r_3) + \bar\psi(\vec r_4)
  \right] \,.
\end{equation}
Finally, if we use the linear relations for the $e_0$ and $e_1$ currents instead of $\bar e_0$ and $\bar e_1$, we find the second part of the Dirac equations, $\partial \bar\psi = im \ \psi$.




\section{Conclusions}
\label{sec:conclu}

We have constructed the quasi-local operators associated to the $\uq$ symmetry underlying the chiral Potts model, for any choice of integrable Boltzmann weights. The half-currents associated with these operators, when dressed with suitable local phase factors, satisfy ``twisted'' discrete Cauchy-Riemann equations~\eqref{eq:contour} of the form
$$
e^{i\phi_r/N} \delta z_1\ \cO(\vec r_1) + e^{i\phi_s/N} \delta z_2\ \cO(\vec r_2)
e^{-i\phi_r/N} \delta z_3\ \cO(\vec r_3) + e^{-i\phi_s/N} \delta z_4\ \cO(\vec r_4) =0\,, 
$$
where $\phi_r$ and $\phi_s$ are the functions \eqref{eq:param} of the spectral parameters $r$ and $s$ along the two directions of the lattice. At the isotropic critical point (FZ clock model), we have exhibited the algebraic origin of the lattice $\Z_N$-parafermions of~\cite{RajCardy}. In the generic case $N \geq 3$, and in the vicinity of the FZ point, we have shown that the above equation actually encodes (a discrete version of) the modified current conservation relation induced by a chiral perturbation of the $\Z_N$-parafermionic CFT. In the Ising case, this equation also allows us to recover the discrete massive Dirac equation of~\cite{RivaCardy}.

Thus, in the framework of the chiral Potts model, we have shown that the quantum group symmetry can be exploited to construct off-critical  discrete parafermions and to probe the nature of the underlying perturbed conformal field theory.

\subsection*{Acknowledgments} The authors would like to thank John Cardy and Paul Zinn-Justin for useful discussions, and to acknowledge and thank Paul Fendley for explaining his published and unpublished work on parafermions  \cite{Fendley13,Fendley14}.

\bibliography{dh.bib}

\end{document}